\documentclass[aps, prx, a4paper, reprint, superscriptaddress, floatfix]{revtex4-2}

\usepackage{amsmath, amssymb}
\usepackage{color, hyperref}
\usepackage[all]{hypcap} 
\usepackage{graphicx}
\usepackage{xcolor}
\usepackage{placeins}
\usepackage[normalem]{ulem}
\usepackage{bm}

\definecolor{MyDarkGreen}{rgb}{0,0.6,0}
\definecolor{MyDarkBlue}{rgb}{0,0,0.8}
\definecolor{MyDarkRed}{rgb}{0.6,0,0.3}
\hypersetup{breaklinks=true,colorlinks=true,plainpages=true,linktocpage=true,linkcolor=MyDarkBlue,citecolor=MyDarkGreen,urlcolor=MyDarkRed,pdfborder={0 0 0},%
pdfauthor={Rene Wagner},%
pdftitle={Helium paper}%
}

\newlength{\figurewidth}
\newcommand{\etal}{{\it et al.}}
\begin{document}

\title{
Circular Dichroism in Multi\-photon Ionization of Resonantly Excited Helium Ions near Channel Closing
}

\author{Ren\'e Wagner}
\affiliation{European X-Ray Free-Electron Laser Facility, 22869 Schenefeld, Germany}
\affiliation{Department of Physics, Universit\"at Hamburg, 22607 Hamburg, Germany}

\author{Markus Ilchen}
\affiliation{Department of Physics, Universit\"at Hamburg, 22607 Hamburg, Germany}
\affiliation{Deutsches Elektronen-Synchrotron DESY, Notkestr. 85, 22607 Hamburg, Germany}
\affiliation{European X-Ray Free-Electron Laser Facility, 22869 Schenefeld, Germany}
\affiliation{Institut f{\"u}r Physik und CINSaT, Universit{\"a}t Kassel, 34132 Kassel, Germany}

\author{Nicolas Douguet}
\affiliation{Department of Physics, University of Central Florida, Orlando, Florida 32816, USA}

\author{Philipp Schmidt}
\affiliation{European X-Ray Free-Electron Laser Facility, 22869 Schenefeld, Germany}

\author{Niclas Wieland}
\affiliation{Department of Physics, Universit\"at Hamburg, 22607 Hamburg, Germany}

\author{Carlo Callegari}
\affiliation{Elettra-Sincrotrone Trieste S.C.p.A., 34149 Basovizza, Trieste, Italy}

\author{Zachary Delk}
\affiliation{Department of Physics, Kennesaw State University, Marietta, Georgia 30060, USA}

\author{Alexander Demidovich}
\affiliation{Elettra-Sincrotrone Trieste S.C.p.A., 34149 Basovizza, Trieste, Italy}

\author{Michele Di Fraia}
\affiliation{Elettra-Sincrotrone Trieste S.C.p.A., 34149 Basovizza, Trieste, Italy}

\author{Jiri Hofbrucker}
\affiliation{Helmholtz-Institut Jena, Fröbelstieg 3, D-07743 Jena, Germany}

\author{Michele Manfredda}
\affiliation{Elettra-Sincrotrone Trieste S.C.p.A., 34149 Basovizza, Trieste, Italy}

\author{Valerija Music}
\affiliation{European X-Ray Free-Electron Laser Facility, 22869 Schenefeld, Germany}
\affiliation{Institut f{\"u}r Physik und CINSaT, Universit{\"a}t Kassel, 34132 Kassel, Germany}
\affiliation{Deutsches Elektronen-Synchrotron DESY, Notkestr. 85, 22607 Hamburg, Germany}

\author{Oksana Plekan}
\affiliation{Elettra-Sincrotrone Trieste S.C.p.A., 34149 Basovizza, Trieste, Italy}

\author{Kevin C. Prince}
\affiliation{Elettra-Sincrotrone Trieste S.C.p.A., 34149 Basovizza, Trieste, Italy}

\author{Daniel E. Rivas}
\affiliation{European X-Ray Free-Electron Laser Facility, 22869 Schenefeld, Germany}

\author{Marco Zangrando}
\affiliation{Elettra-Sincrotrone Trieste S.C.p.A., 34149 Basovizza, Trieste, Italy}
\affiliation{CNR Istituto Officina dei Materiali, Laboratorio TASC, 34149 Basovizza, Trieste, Italy}

\author{Alexei N. Grum-Grzhimailo}

\affiliation{Kirovogradskaya 40-2-216, 117534 Moscow, Russia}

\author{Klaus Bartschat}
\affiliation{Department of Physics and Astronomy, Drake University, Des Moines, Iowa 50311, USA}

\author{Michael Meyer}
\affiliation{European X-Ray Free-Electron Laser Facility, 22869 Schenefeld, Germany}

\begin{abstract}
The circular dichroism (CD) of photo\-electrons generated by near-infra\-red (NIR) laser pulses using multi\-photon ionization of excited He$^+$ ions in the \hbox{$3p\,(m\!=\!+1)$} state. The ions were prepared by circularly polarized extreme ultra\-violet (XUV) pulses.  For circularly polarized NIR pulses co- and counter-rotating relative to the polarization of the XUV pulse, a complex variation of the CD is observed as a result of intensity- and polarization-dependent Freeman resonances, with and without additional dichroic AC-Stark shifts. The experimental results are compared with numerical solutions of the time-dependent Schr\"odinger equation to identify and interpret the pronounced variation of the experimentally observed CD.
\end{abstract} 

\maketitle

\section{Introduction}

Photo\-ionization with circularly polarized light can provide unique information about the dynamics and fundamentals of light-matter interaction. 
In particular, the circular dichroism (CD) in photo\-emission
experiments, i.e., the difference in the ionization signal between configurations using opposite light helicities, were successfully explored at synchrotron radiation facilities \cite{baier1994,Aloise2005} and with intense optical lasers \cite{loge2011multiphoton, eckart2018ultrafast, de2021using}. Various phenomena can be examined, such as isotope effects in  atomic photo\-ionization, which reveal the coupling between electronic and nuclear angular momenta \cite{gryzlova2015}, information on the chirality of molecular systems, e.g., via measurements of the photoelectron circular dichroism \cite{ritchie1976theory, bowering2001asymmetry, nahon2015, beaulieu2018, ilchen2021site} and its relevance for ionization and dissociation processes \cite{veyrinas2019}, as well as the magnetic properties and magnetization dynamics of various materials \cite{van2014}. In studies dedicated to tunnel ionization via strong fields \cite{Barth11,eckart2018ultrafast}, it was predicted and demonstrated that ionization is favored for electrons that are initially orbiting against the laser field. Whether this is a general phenomenon or only evident under specific conditions is still under debate. In pump-probe experiments, ionization with counter-rotating pulses could thus be stronger than ionization with co-rotating pulses.

Recent experiments dedicated to CD studies in the nonlinear regime \cite{Mazza2014,kazansky2011, Ilchen17} have been made possible by short-wavelength free-electron lasers (FELs). One commonly used tool is the so-called ``sideband method'', where the photo\-electrons are dressed by a synchronized optical or infrared (IR) laser field. This results in additional spectral contributions to the photo\-electron line, which are separated by the energy of the optical laser photons. The variation of the relative intensity of the sidebands for different combinations of circularly polarized FEL and optical pulses was used to determine the polarization state of \hbox{X-ray} or XUV radiation \cite{Mazza2014, hartmann2016circular} and the relative intensity of partial-wave contributions~\cite{mazza2016}. 

Even more information about the complex interactions in the electronic cloud is accessible via the study of resonances and transient states \cite{rorig2023multiple}. The precise characterization of their energy position, absorption strength, and spectral profile provides rich information about both the interactions in the electronic sub\-shells of the target atom and the coupling to the ionization continuum. In particular, high-intensity laser fields can be used to steer specific and also transient states into resonance, a phenomenon known as Freeman resonances \cite{freeman1991,marchenko2010}. In this case the photo\-ionization dynamics are strongly affected by the symmetry of the intermediate resonant state. 
 
In order to study the fundamental principles in detail experimentally and theoretically, small systems with few or preferably only one electron are desirable. In this regard, single-electron helium ions are ideal targets, since the alternative, the hydrogen atom, is particularly challenging to investigate experimentally. Hydrogen-like oriented helium ions created through sequential ionization and absorption by a circularly polarized FEL pulse thus offer a vital alternative for detailed investigations of dichroic phenomena \cite{Ilchen17} and the respective role of inter\-mediate resonances. Studies of the latter have been elusive in practice until very recently, in particular regarding their role in the photo\-ionization of transient and ionic systems.

\begin{figure}[!t]
\includegraphics[width=0.620\columnwidth]{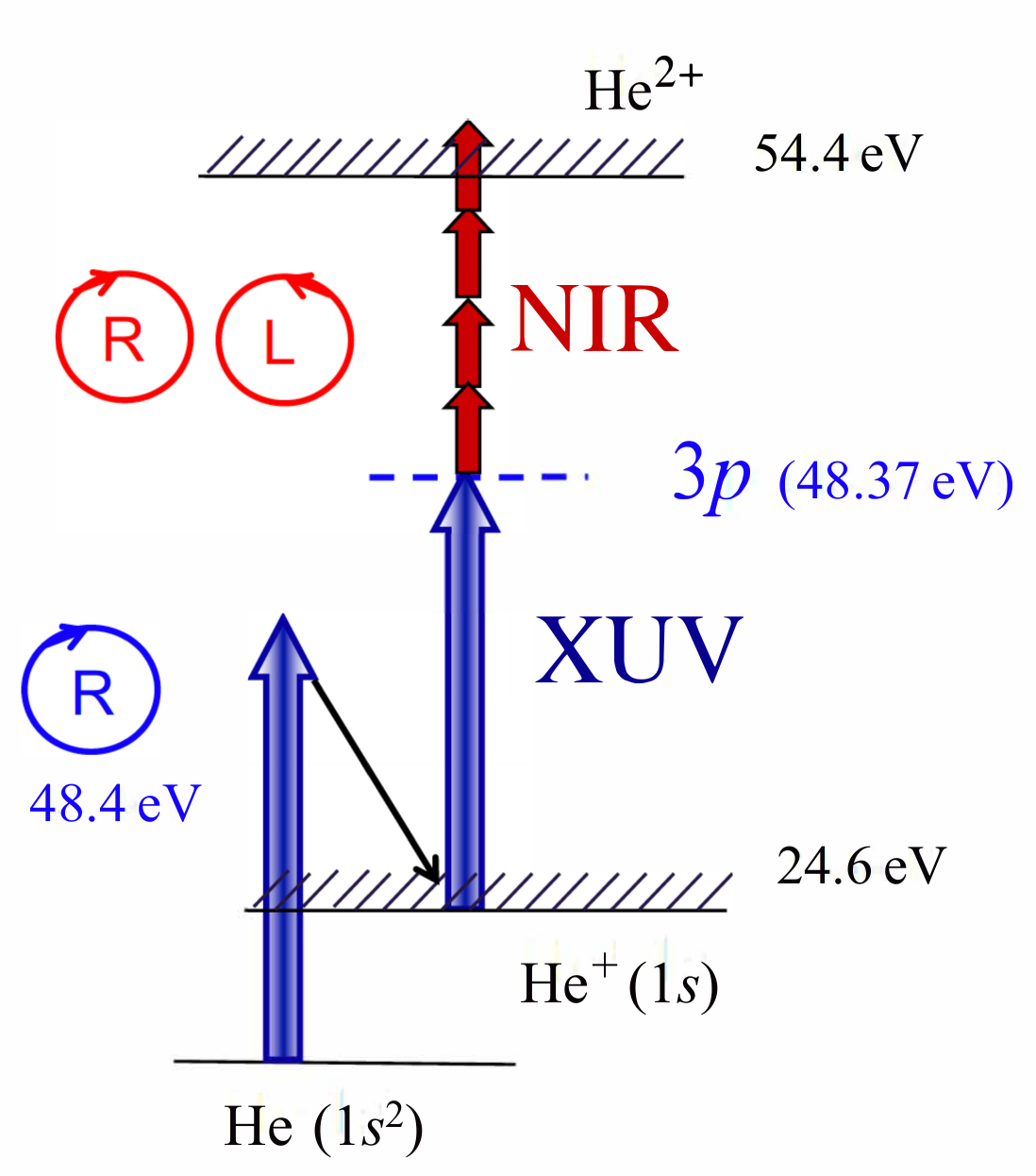}
\label{fig:He_3p_scheme}
\caption{Schematic representation of the ionization and excitation processes. Right-handed circularly polarized XUV radiation ionizes the helium atom and subsequently excites the He$^+(3p,m\!=\!+1)$ resonance. Intense left- or right-handed circularly polarized NIR pulses further ionize the excited ions via a multi\-photon process.
The energies of the $3p$ state and the ionization threshold of He$^{2+}$ are given relative to the He$^+(1s)$ ground state.}
\end{figure}

In the present work, laser-based multi\-photon ionization (MPI) via a \hbox{He$^{+}3p\,(m\!=\!+1)$} oriented state created by intense, circularly polarized, extreme ultra\-violet (XUV) radiation from the XUV-FEL FERMI~\cite{fermi_allaria} is used to study the polarization- and intensity-dependent yields, kinetic energies, and dichroic properties of the resulting photo\-electrons (see Fig.~\ref{fig:He_3p_scheme}). In particular, we investigate the dependence of the CD over a range of near-infra\-red (NIR) laser intensities ranging from 10$^{12}$\,W/cm$^2$ to 10$^{13}$\,W/cm$^2$ with a central wavelength of 784~nm. The CD is defined as 
\begin{equation}
{\rm CD} \equiv \frac{P_{+} - P_{-}}{P_{+}\!+\!P_{-}}.
\end{equation}
Here $P_{+}$ and $P_{-}$ are the probabilities for
ionization by circularly polarized pulses with the same~($+$) or opposite~($-$) helicity, respectively.

In our earlier work \cite{Ilchen17}, the strong influence of the AC-Stark shift on the intensity dependence of the CD was explored. 
In the present study, we eliminate this influence on the population of the He$^{+}$ ions in the \hbox{$3p\,(m\!=\!+1)$} oriented state by delaying the optical laser pulse by 500~fs in time, i.e., we first create the oriented ions and subsequently ionize them with the optical laser after the FEL pulse has fully passed. 

The chosen intensity range includes not only multi-photon interactions but also over-the-barrier effects and can thus yield information about characteristic regimes of the Keldysh parameter in order to draw a  comprehensive picture of circular dichroism in a wide range of field strengths. Since for higher NIR intensities the binding energy of electrons is being shifted to values higher than four times the NIR photon energy, the electrons require an additional NIR photon to enter the continuum. 
In the corresponding intensity range, we reveal not only the intensity-dependent role of \hbox{Rydberg} resonances on the dichroic yields of multi\-photon ionization (MPI) and above-threshold ionization (ATI), but also analyze the previously debated sign change of the dichroism that was hitherto only predicted by theory for an AC-Stark-shift-influenced system \cite{Ilchen17}.
Our observations thus contribute to shedding light on fundamental properties of dichroic light-matter interactions and the claimed predominant role of counter-rotating fields in the high-intensity regime \cite{Barth11, Bauer14, eckart2018ultrafast}. 

This paper is organized as follows.  Section~\ref{sec:Methods} summarizes the experimental (\ref{subsec:exp}) and theoretical (\ref{subsec:theory}) methods employed in the present work. This is followed by a presentation and discussion of our results in Sect.~\ref{sec:Results}.  Our conclusions and an outlook are given in Sect.~\ref{sec:conclusion}.

\section{Methods}\label{sec:Methods}
\subsection{Experiment}
\label{subsec:exp}

The experiments were performed at the {L}ow-{D}ensity {M}atter (LDM) 
end-station~\cite{Lyamayev_2013} of the seeded free-electron laser
FERMI~\cite{allaria2012}. The circularly polarized XUV pulses from 
the FEL-1 laser were tuned to the 
$3p$ resonance of the helium ion at 48.37\,eV (25.63\,nm)~\cite{NIST} (under field-free conditions), which was excited by sequential two-photon absorption within the FEL pulse duration of 75\,fs\,$\pm$ 25\,fs (FWHM).
 
With pulse energies up to 150\,µJ and a focus size of \hbox{10-20\,µm} (FWHM), the peak intensity of the XUV radiation was about $10^{13}$\,W/cm$^2$. 
The repetition rate of the FEL pulses was 50\,Hz, and the degree of circular
polarization was 95\,$\%$ $\pm$ 5\,$\%$~\cite{Mazza2014}. 

The circularly polarized NIR laser had a mean photon energy of 1.58\,eV (784\,nm), 
a bandwidth (FWHM) of 26\,meV (13\,nm), and an ellipticity of $\approx\,$0.4. 
It is noteworthy that, in comparison with a perfectly circular polarization of the NIR, our numerical simulations reveal only a minor reduction of about $10\,\%$ in the expected CD.
With a pulse duration of 140\,fs $\pm$ 30\,fs (FWHM), a focus spot size of 150\,µm $\pm$ 50\,µm (FWHM),
and pulse energies ranging from 30\,µJ to 420\,µJ, 
laser intensities from $0.9 \times 10^{12}$\,W/cm$^2$ to 
$1.1 \times 10^{13}$\,W/cm$^2$ were available. 

The helium gas was injected into the experimental chamber (background pressure $\approx\,6.5\times 10^{-9}$\,mbar) via super\-sonic expansion. In the center of a velocity map imaging (VMI) 
electron spectro\-meter~\cite{Lyamayev_2013}, the gas jet crossed the spatially
overlapping, co-propagating XUV and NIR beams. The energy resolution of the VMI spectrometer was 90\,meV (FWHM) at a kinetic energy of 0.1\,eV and better than 200\,meV over the entire range of the present investigation. Due to the applied setting of the VMI, electrons with kinetic energies of less than about 50\,meV could not be reliably detected in our experiment, and hence their signal is not reported here.

\subsection{Theory}
\label{subsec:theory}
As indicated in the introduction, we consider the case without an AC Stark shift in the preparation process of the ions. The XUV pulse first ionizes one of the $1s$ electrons in the helium ground state and subsequently excites the other one to the $3p$ level. After a delay of about 500~fs, which is very small compared to the lifetime of $\approx750\,$ps of the He$^+(3p)$ excited state~\cite{NIST},
the relaxation of the $1s$ orbital from
the neutral to the ionic configuration has taken place already, while no significant radiative decay
of the excited ionic state has happened. 
We therefore solve the time-dependent Schr\"odinger equation (TDSE) for the active electron, which is the remaining bound electron after the initial ionization step of the He$(1s^2)$ ground state \cite{Ilchen17}. 

Consequently, we are dealing
with a true one-electron target, whose non\-relativistic 
orbitals are known analytically~\cite{bethe2013quantum}, while the continuum states on which to project the final-state
wave function
to obtain the ionization signal are pure Coulomb functions.
Hence, we can start the calculation directly with a mixture of He$^+$ ions in
the $1s$ ground state and the polarized
\hbox{He$^+(3p,m\!=\!+1)$} state, from which the electron
is removed by the co- or counter-rotating circularly polarized 
NIR field. For the parameters in the current experiment, the 
actual mixture of ions in the excited state and the ground state is irrelevant, since
the NIR alone is too weak to seriously affect the $1s$ electron.

Since reporting our earlier work~\cite{Ilchen17}, 
we significantly enhanced
the efficiency of the associated computer code through improved
OpenMP parallelization and a variable radial grid in our finite-difference
method~\cite{atoms12070034}. Specifically, we take a smallest stepsize of 0.01 
(we use atomic units in this section)
near the nucleus and a largest stepsize of 0.05 for large 
distances. All these
improvements made it possible to apply pulses of duration and peak intensity much closer to those used in the experiment, which could not be done previously~\cite{Ilchen17}.

The calculations were carried out in the velocity 
gauge of the electric dipole operator. It is well 
known~\cite{CorLam1996, Grum10}
that partial-wave convergence for long wavelengths is much faster 
in this gauge than in the length gauge. This aspect is crucial for the present 
problem, since we cannot use the simplifying cylindrical symmetry of linearly
polarized light.  Specifically, partial waves up to angular momenta~$\ell = 40$ were included in order to ensure converged results.
Tests carried out by varying the time step, the radial grid, 
and the number of partial waves, as well as their magnetic quantum numbers, 
give us confidence in the numerical accuracy of the predictions. Most likely, the
largest uncertainty originates from the fact that we  
make idealized assumptions about the experimental pulse, such as its length, shape, peak intensity, polarization, and carrier envelope phase, i.e., parameters that are generally difficult to determine accurately in the experiment.

In the present calculations, we took a 160-cycle pulse of the NIR laser
(60\,fs FWHM in intensity) with a sine-square 
envelope of the vector potential~$\bm{A}(t)$, from which we calculated the
electric field as $\bm{E}(t) = -\partial\bm{A}(t)/\partial t$.  Setting the
vector potential rather than the electric field directly avoids a possibly unphysical 
pulse with non\-vanishing displacement~\cite{PhysRevA.90.043401}, 
although for long
pulses like those employed here the effect of differentiating the envelope function
is small. Furthermore, changing the carrier-envelope phase is not important
for such pulses either.  Finally, the target created by the XUV pulse 
is very small compared to the spot irradiated by the NIR pulse.  Hence, focal-point
averaging of the NIR intensity over the interaction volume is not necessary either.

\section{Results and Discussion}\label{sec:Results}

In order to validate the approach of using temporally separated XUV and NIR pulses to eliminate the contribution of the AC-Stark shift from the present observation, we first reproduced the conditions of the earlier experiment~\cite{Ilchen17} performed with temporally overlapping pulses. Under these conditions the observed intensity-dependent changes of the CD are mainly caused by the polarization-dependent AC-Stark shift of the $3p$ resonance~\cite{PhysRevA.100.033404}. To further test this interpretation, electron spectra were recorded while scanning the XUV pulses around the He$^+(3p)$ resonance by varying the photon energy from 48.15\,eV to 48.60\,eV in the presence of the temporally overlapping optical laser. 

\begin{figure}[!t]
\includegraphics[width=0.47\textwidth]{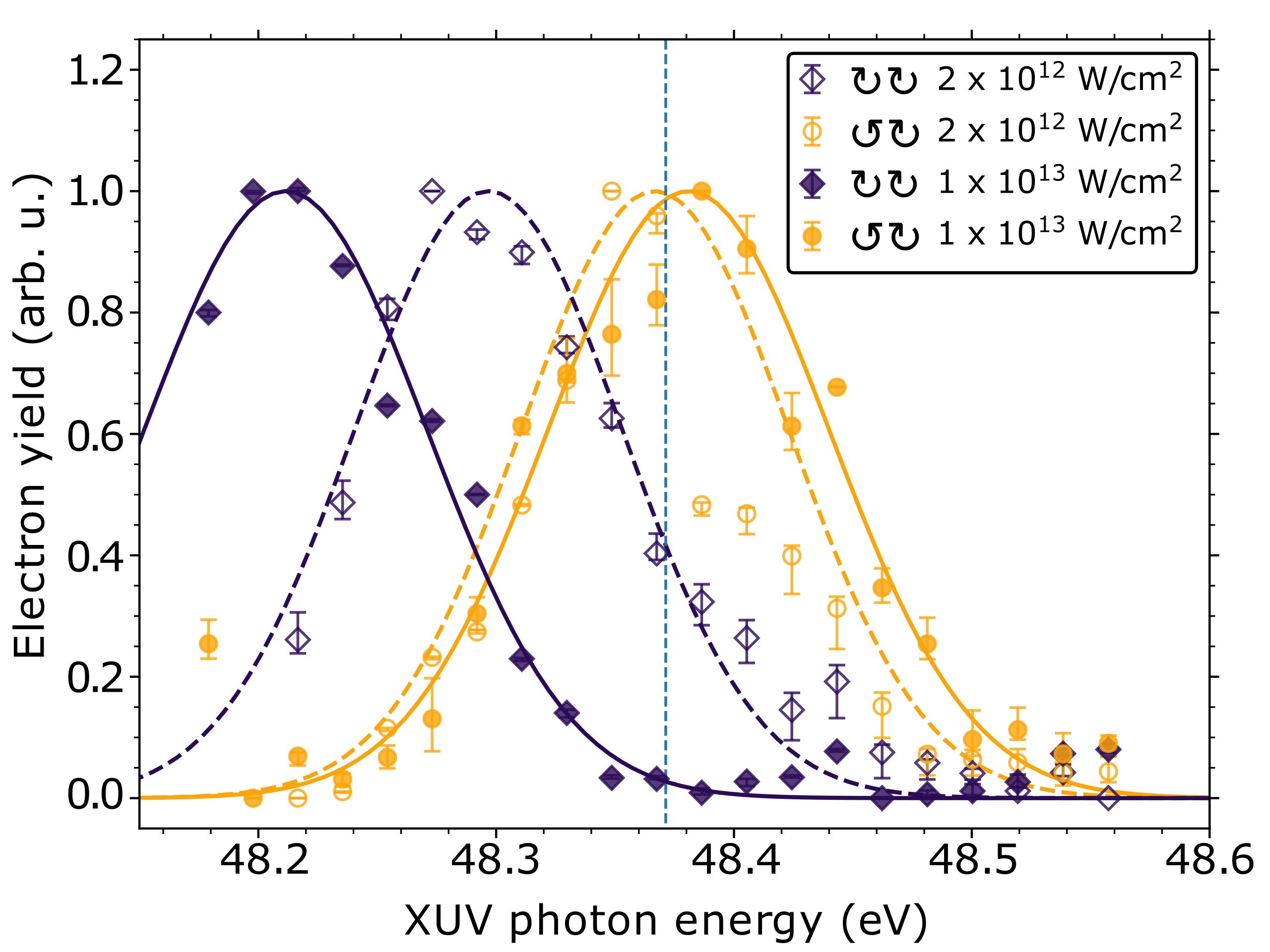}
\label{fig:3p_resonance_shift}
\caption{Photo\-electron yield generated by co-  or counter-rotating polarized 
optical pulses depending on the XUV photon energy at NIR-laser intensities of $2\times10^{12}$\,W/cm$^2$ and $1\times10^{13}$\,W/cm$^2$. 
The dashed vertical line marks the position of the $3p$ resonance in the absence of the 
optical laser. 
The peaks for the co-rotating cases are shifted significantly to the left of the field-free line.
Gaussian profiles (lines) were fitted to the experimental data points 
to highlight the polarization- and intensity-dependent shifts.}
\end{figure} 

The yields of the MPI and ATI photo\-electrons, produced upon multi\-photon ionization of the $3p$ resonance by the NIR laser, are displayed in Fig.~\ref{fig:3p_resonance_shift} as a function of the FEL photon energy for two NIR peak intensities with co- and counter-rotating circular laser polarization, respectively. As demonstrated by the experimental data and discussed in detail by 
Grum-Grzhimailo \etal~\cite{PhysRevA.100.033404}, the shift of the resonance position indeed strongly 
depends on both the relative helicity and the laser intensity. 
While for counter-rotating polarization the position coincides within the experimental resolution with the field-free position at 48.37\,eV, the data recorded for co-rotating polarization clearly show a shift of the resonance, which grows with increasing NIR-laser intensities. The measured energy shift of the resonance by 76\,meV towards lower excitation energies for a laser intensity of $1\times10^{12}$\,W/cm$^2$ is in good agreement with the theoretical value of 90\,meV. 
 For the counter-rotating case, a shift towards higher energies of no more than 
10\,meV up to a laser intensity of $6\times10^{12}$\,W/cm$^2$ was calculated. 
Hence, if the pulses overlap as in our earlier
work~\cite{Ilchen17}, the observed dichroism is indeed strongly affected
by how many ions are actually prepared in the oriented excited state under investigation. The population of this excited state, therefore, is one of the crucial parameters in determining the CD.

In order to eliminate the effect of this AC-Stark-shift-induced ``preparation asymmetry'' of the oriented ions, all further experiments were performed with the NIR pulses delayed by about 500~fs relative to the XUV. This makes it possible to analyze in detail the influence of the intense optical laser field on the ionization probabilities for different combinations of the circularly polarized XUV and NIR pulses, thus giving access to studying the Freeman resonances in better detail.

\begin{figure}[!t]
\includegraphics[width=0.47\textwidth]{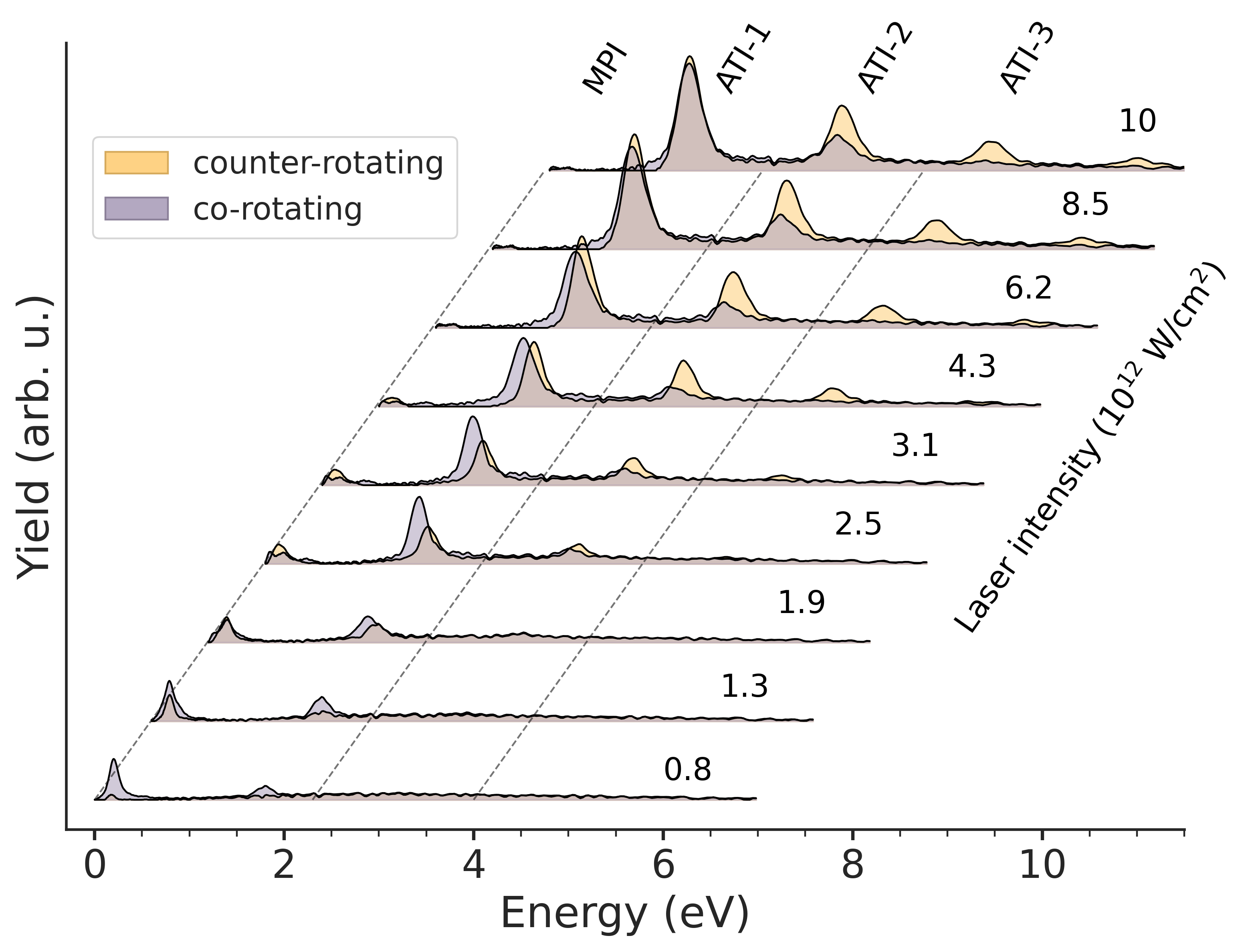}
\label{fig:CD_RadialProfile}
\caption{Photo\-electron spectra for different peak intensities of the optical laser for co- (darker shade) and counter-rotating (lighter shade) polarized NIR laser pulses.
The zero lines for each peak intensity (indicated in multiples of 10$^{12}\,$W/cm$^2$
on the right) have been offset to make the individual results visible.}
\end{figure}

Figure~\ref{fig:CD_RadialProfile} exhibits photo\-electron spectra for a range
of peak intensities of the optical laser for co- and counter-rotating circularly polarized pulses. 
We see strong intensity variations in the peak at the lowest kinetic energies (around 0.2\,eV). At low intensities, the signal for co-rotating pulses stays nearly constant, while the signal is strongly increasing for counter-rotating pulses. As predicted in~\cite{PhysRevA.100.033404}, 
the MPI peak, which originates from the absorption of four NIR photons, quickly moves below 
threshold. 
The signal in the counter-rotating 
case remains almost constant up to an intensity of about $4.3 \times 10^{12}\,$W/cm$^2$. This behavior results in the general trend of the CD observed already earlier, i.e., a maximum circular dichroism
of $\approx\,+1$ at low NIR intensity, followed by a successive drop with increasing intensity (see Fig.~\ref{fig:CD_allFeatures} below).
Figure~\ref{fig:CD_RadialProfile} also clearly shows the first ATI peak (labeled \hbox{ATI-1}) at all displayed intensities, followed by the second (\hbox{ATI-2}) starting to be visible around 
\hbox{$2 \times 10^{12}\,$W/cm$^2$}, and even a third (ATI-3) at the highest peak intensities.

As seen in Fig.~\ref{fig:CD_allFeatures}, both experiment and theory predict nearly maximum values of the CD for the MPI line at the lowest peak intensity. However, the sign change cannot be confirmed for the case without an AC-Stark shift of the $3p$ resonance in the ion. 
Taking the experimental conditions into account also for the theory by cutting off the signal below 50~meV, a sign change around
$2 \times 10^{12}\,$W/cm$^2$ and another sign change back to positive
values around $5 \times 10^{12}\,$W/cm$^2$ are consistently found in both experiment and theory. The remaining differences are predominantly originated by the experimental intensity uncertainty. 
However, the kinetic energy range from 0 to 50 meV is calculated to yield significant contributions to the CD of the MPI (ocher circles in Fig.~\ref{fig:CD_allFeatures}). Theory thus predicts the CD of the MPI peak to stay positive throughout the intensity range of the present study. 

We emphasize that in the current setup with non-overlapping pulses, the CD value is {\it not\/} affected by less ions in the initial state due to the NIR acting already during the preparation stage. 
Consequently, the AC-Stark shift, which was predicted to lead to negative CD values for the MPI, is strongly influencing the CD, but it is not the only process resulting in significant CD variations at low NIR intensities. Other phenomena have to be taken into account to explain the observations under the present conditions. 

\begin{figure}[!t]
\includegraphics[width=0.47\textwidth]{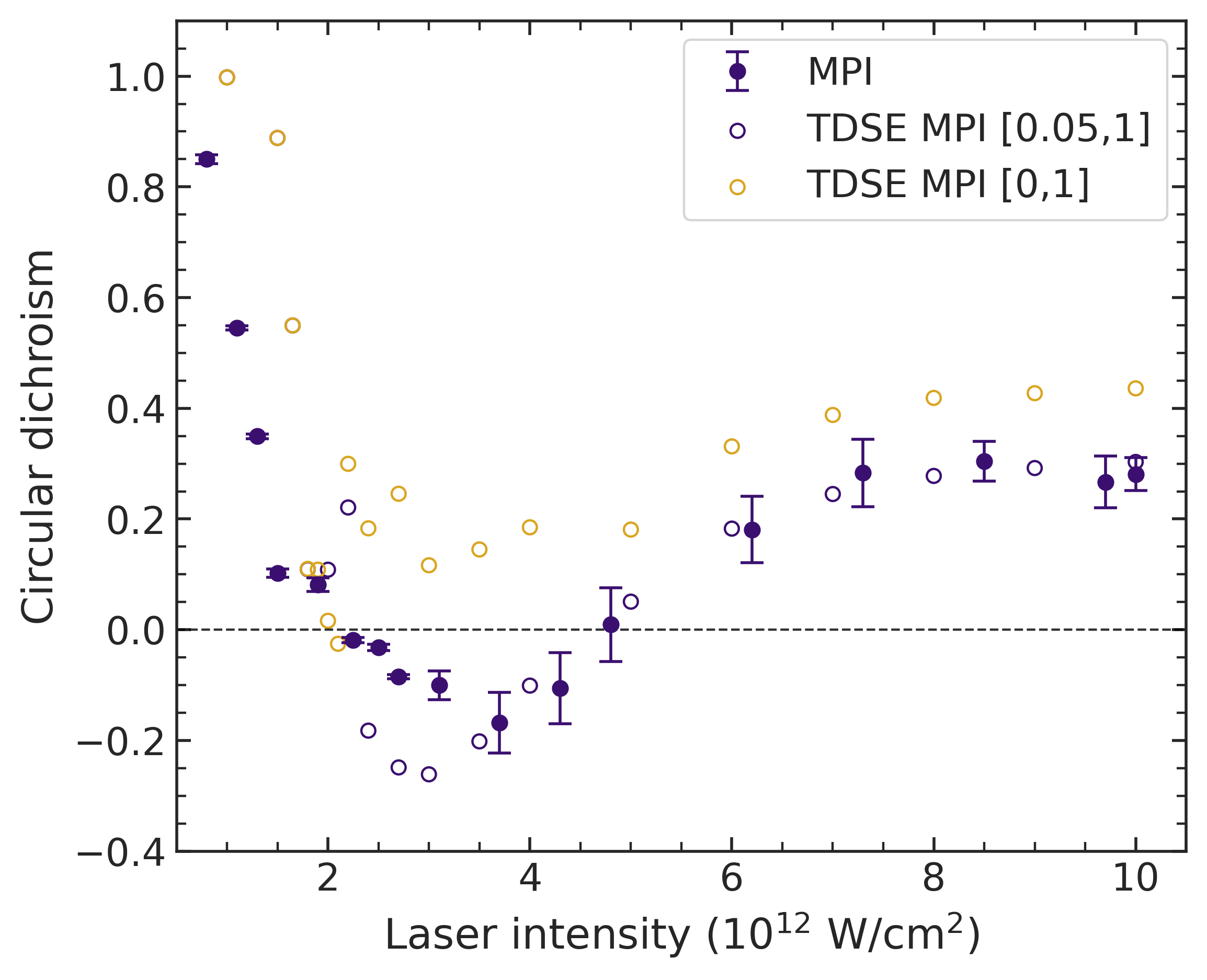}
\caption{Experimental and theoretical CD values at different NIR-laser peak intensities for the MPI feature of the photo\-electron spectrum. 
The solid symbols with error bars represent the experimental data, 
while the TDSE predictions are shown as open symbols. The theoretical predictions are displayed once as adapted to the experimental conditions (blue) and again including all features down to zero kinetic energy (ocher). The numbers in parentheses of the legend indicate the kinetic energy range (in eV) that was accounted for in the integration over the theoretical MPI peak.
}
\label{fig:CD_allFeatures}
\end{figure}

The general trend of the decreasing CD and the sign change for the MPI line can be explained by the complex interplay between the shift of the MPI line and channel closing, which is different in the co- and counter-rotating cases. While direct ionization by a four-photon process is no longer possible for intensities higher than 
$2.5 \times 10^{12}\,$W/cm$^2$ in the case of co-rotating pulses, channel closing appears only at $4.8 \times 10^{12}\,$W/cm$^2$ for counter-rotating pulses, as calculated in~\cite{PhysRevA.100.033404}.

\begin{figure}[!t]
\includegraphics[width=0.47\textwidth]{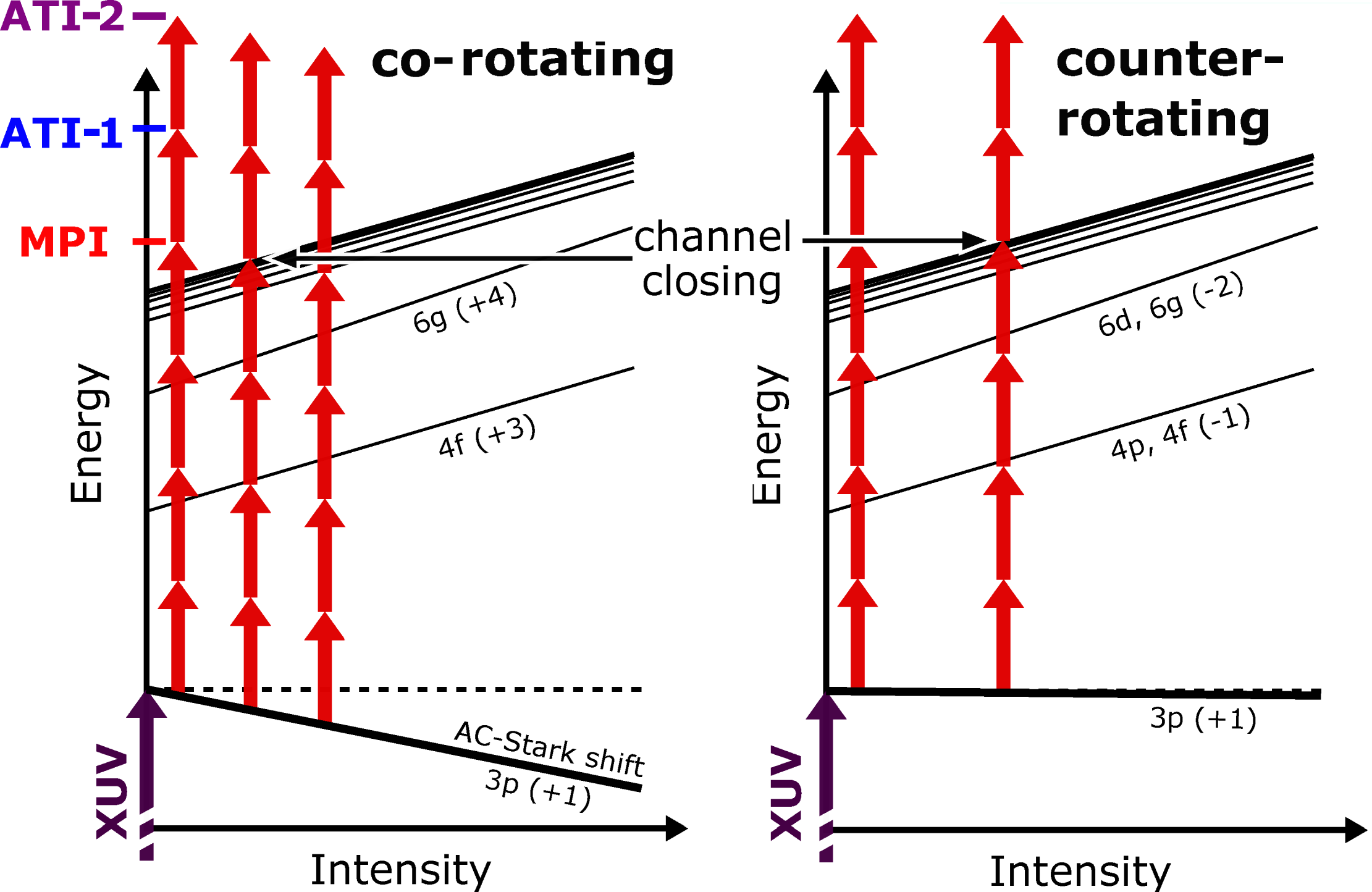}
\label{fig:helium_levels}
\caption{Schematic illustration of multi\-photon ionization from the He$^+(3p)$ level 
for co- and counter-rotating polarized XUV and NIR pulses.  The 
arrows
represent the NIR and XUV photons, respectively. 
The tilted lines indicate various states. For a detailed discussion of 
the individual shifts and the most important intermediate resonance states involved,
see~\cite{PhysRevA.100.033404}. Note that the four-photon
channel, which leads to the main photoline (MPI), closes at significantly lower NIR intensity for the co-rotating (about $2.5 \times 10^{12}\,$W/cm$^2$) compared 
to the counter-rotating case (about $4.8 \times 10^{12}\,$W/cm$^2$).  Also indicated are the first two above-threshold 
ionization lines \hbox{ATI-1} and \hbox{ATI-2}.
}
\end{figure}

A simplified general structure of the energy levels in He$^+$ in the presence of an intense external field is illustrated 
for the current situation in Fig.~\ref{fig:helium_levels}, i.e., after preparing the excited He$^+(3p)$ state with right-handed circularly polarized XUV radiation.
The peak position of the $3p$ resonance hardly changes with increasing NIR intensity for the counter-rotating case, whereas it is shifted significantly
towards lower excitation energies for the co-rotating case. This strong variation of the $3p$ resonance position has, together with the general shift of the ionization threshold and the high-lying Rydberg states, also an effect on the MPI step. In the low-intensity regime, ionization of the $3p$ resonance can be achieved by a four-photon process resulting in the ejection of electrons with kinetic energies of about 0.2\,eV~\cite{Ilchen17}. For increasing NIR intensities, the shift of the levels is transformed to a shift in the kinetic energy of the MPI electron, which rapidly moves towards (and ultimately below) the ionization threshold.  

This behavior is very different from other observations of the circular dichroism in the ionization of polarized intermediate states. A multi-photon ionization study of the \hbox{$2p\, (m\!=\!+1)$} state in atomic Li~\cite{acharya2022two}, performed in the intensity regime $0.3 - 1.3 \times 10^{12}\,$W/cm$^2$, does not show an intensity dependence of the CD, but instead a strong wavelength dependence. Furthermore, at higher intensities around $8.5 \times 10^{14}\,$W/cm$^2$, where tunnel ionization is the dominating mechanism, ionization of Ar atoms with co-rotating pulses was found to be more likely than with counter-rotating pulses \cite{eckart2018ultrafast}. In the present investigation with external fields in the order of a few $10^{12}\,$W/cm$^2$, the dynamics are much more complex, and the CD is determined by the interplay of different mechanisms.

\begin{figure}[!t]
 \includegraphics[width=0.48\textwidth]{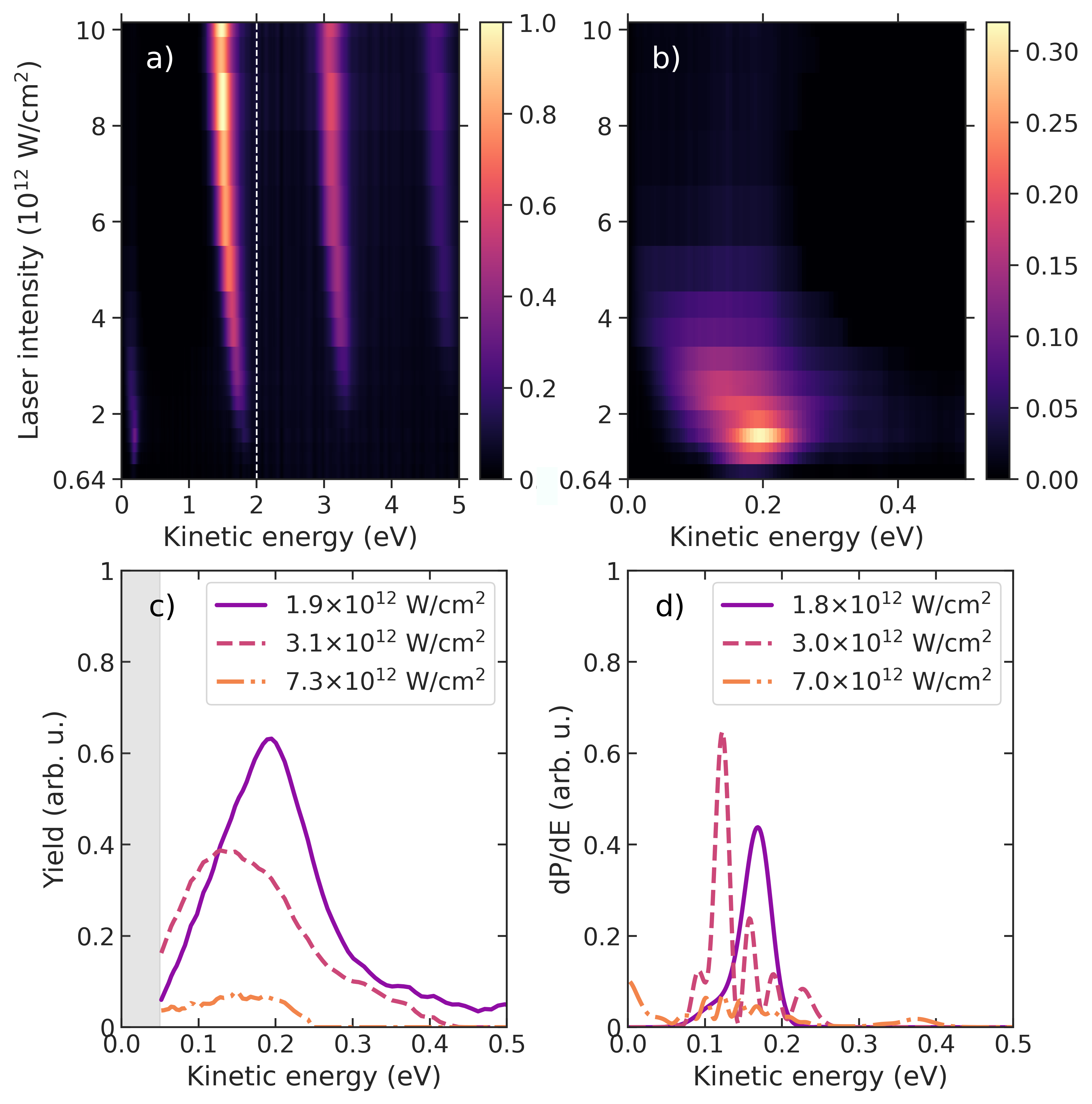}
\label{fig:IntDep_PES_counter}
\caption{Intensity dependence of the photo\-electron spectrum for the counter-rotating case a) for all features and b) focused on the low-energy MPI feature. Note the different scales in the color bars to make the low-energy resonance visible. Panel~c) shows the experimental data as lineouts for the intensities indicated in the legend, with the corresponding theoretical results exhibited in panel~d). The gray-shaded area in c)  indicates the region in the experimental data affected by a lack of transmission and is thus neglected for the CD determination.}
\end{figure} 

\begin{figure}[!t]
\includegraphics[width=0.48\textwidth]{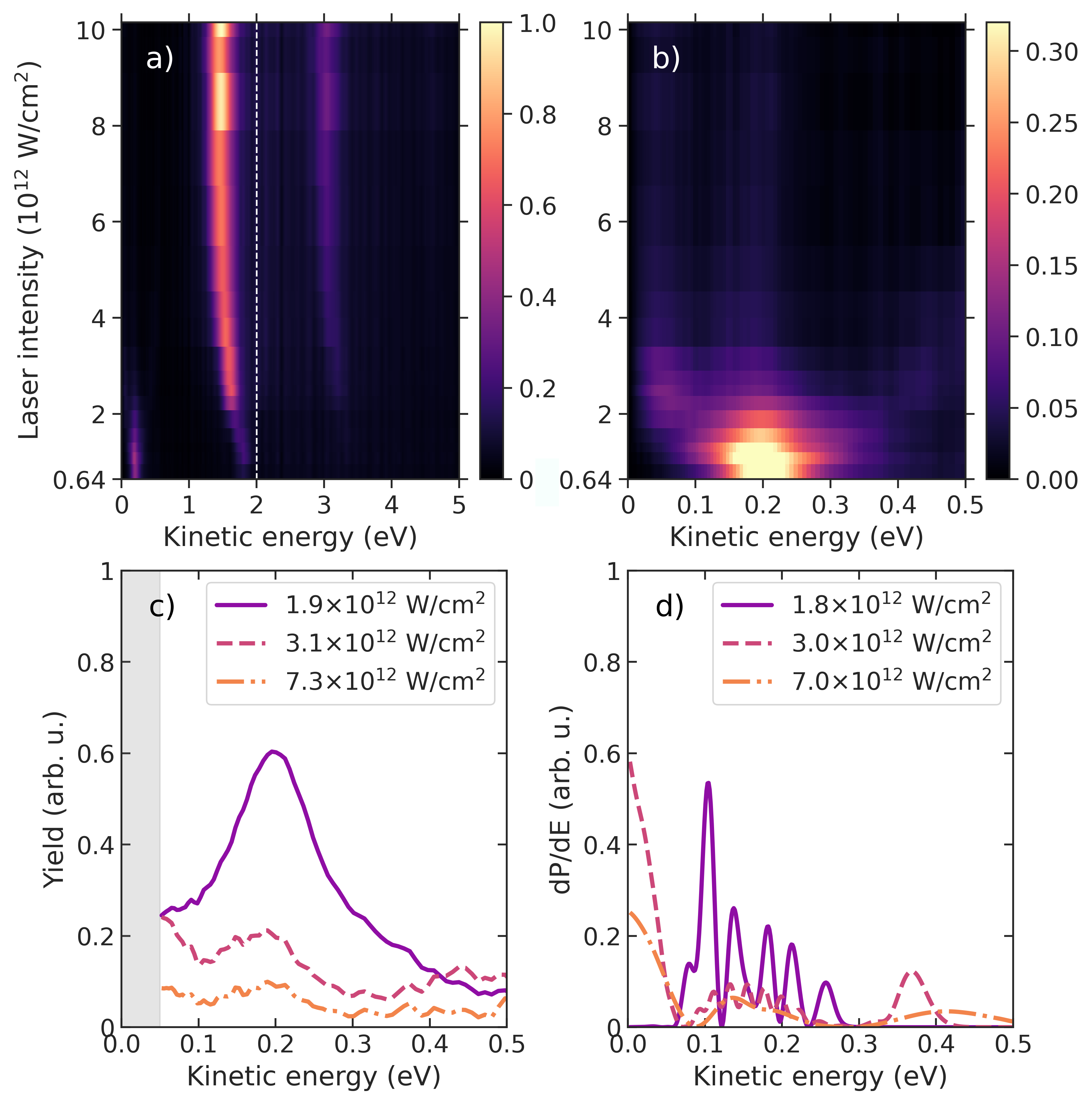}
\label{fig:IntDep_PES_co}
\caption{Same as Fig.~\ref{fig:IntDep_PES_counter} for the co-rotating case. }
\end{figure} 

 Additional insight can be gained by analyzing the individual electron spectra created by multi\-photon ionization at different intensities. Figures~\ref{fig:IntDep_PES_counter} and~\ref{fig:IntDep_PES_co} show these spectra
for the counter- and co-rotating cases, respectively, together
with a few lineouts for selected intensities. The shift of the lines with increasing intensity is nicely seen in the ATI peaks. For the MPI line, which is presented on an enlarged scale in Figs.~\ref{fig:IntDep_PES_counter}(b) and~\ref{fig:IntDep_PES_co}(b), some internal structures besides the shift to lower kinetic energies are also visible and highlighted in the lineouts. For the counter-rotating case (cf.\ Fig.~\ref{fig:IntDep_PES_counter}), a resonance is clearly seen around $1.5 \times 10^{12}\,$W/cm$^2$. These features originate from Freeman resonances~\cite{PhysRevLett.59.1092}, which at particular intensities become resonant for three-photon excitation of the $3p$ state in the four-photon ionization process. Given the strength of the effect under the present conditions, we rule out substantial contributions due to energy shifts induced by nondipole effects, which were recently reported for strong-field and above-threshold ionization \cite{hartung2021electric, lin2022magnetic}. In the present case, these resonances are associated with $6d\,(m\!=\!-2)$ and $6g\,(m\!=\!-2)$, which are clearly separated energetically in theory.  However, they are only visible as a shift in the experimental spectra due to the finite energy resolution of the spectrometer.

For the co-rotating case (cf.\ Fig.~\ref{fig:IntDep_PES_co}), the appearance of the Freeman resonances is even more pronounced, and resonances with symmetry \hbox{$6g\,(m\!=\!4)$} and \hbox{$4f\,(m\!=\!3)$} can be identified around $1.1 \times 10^{12}\,$W/cm$^2$ and $2.5 \times 10^{12}\,$W/cm$^2$. The qualitative agreement
between experiment and theory is good, with some remaining differences in the details.
Interestingly, we note substructures in the theoretical predictions for some, but not
all, of the intensities. These interference structures are particularly pronounced for 
the $6g\,(m\!=\!4)$ resonance in Fig.~\ref{fig:IntDep_PES_co}(d), but are not resolved in the experiment. They arise from inter\-fering wavepackets with the same kinetic energy, but formed at the leading and trailing edges of the pulse~\cite{Wang2020}.
Finally, the persistent signal of the \hbox{$4f\,(m\!=\!+3)$} resonance in the co-rotating case at high NIR intensities is the main reason for re-establishing positive CD values in the MPI line (cf.\ Fig.~\ref{fig:CD_allFeatures}). 

Generally, the influence of the Freeman resonances should also be visible in the high-energy ATI peaks, but the energy resolution of the experiment is limiting its observation.
Besides a strong variation of their intensity ratios, the corresponding lines exhibit an intensity-dependent shift of their line positions, which is different for co- and counter-rotating pulses. The largest difference for \hbox{ATI-1} is observed at $3.1 \times 10^{12}\,$W/cm$^2$ and amounts to 0.11 eV. For \hbox{ATI-2} the difference is most pronounced at $4.1 \times 10^{12}\,$W/cm$^2$ and amounts to 0.15 eV. This behavior is also visible in the numerical simulations and can be interpreted as another manifestation of the excitation of Freeman resonances in the multi-photon ionization pathway. 

\begin{figure}[!t]
\includegraphics[width=0.47\textwidth]{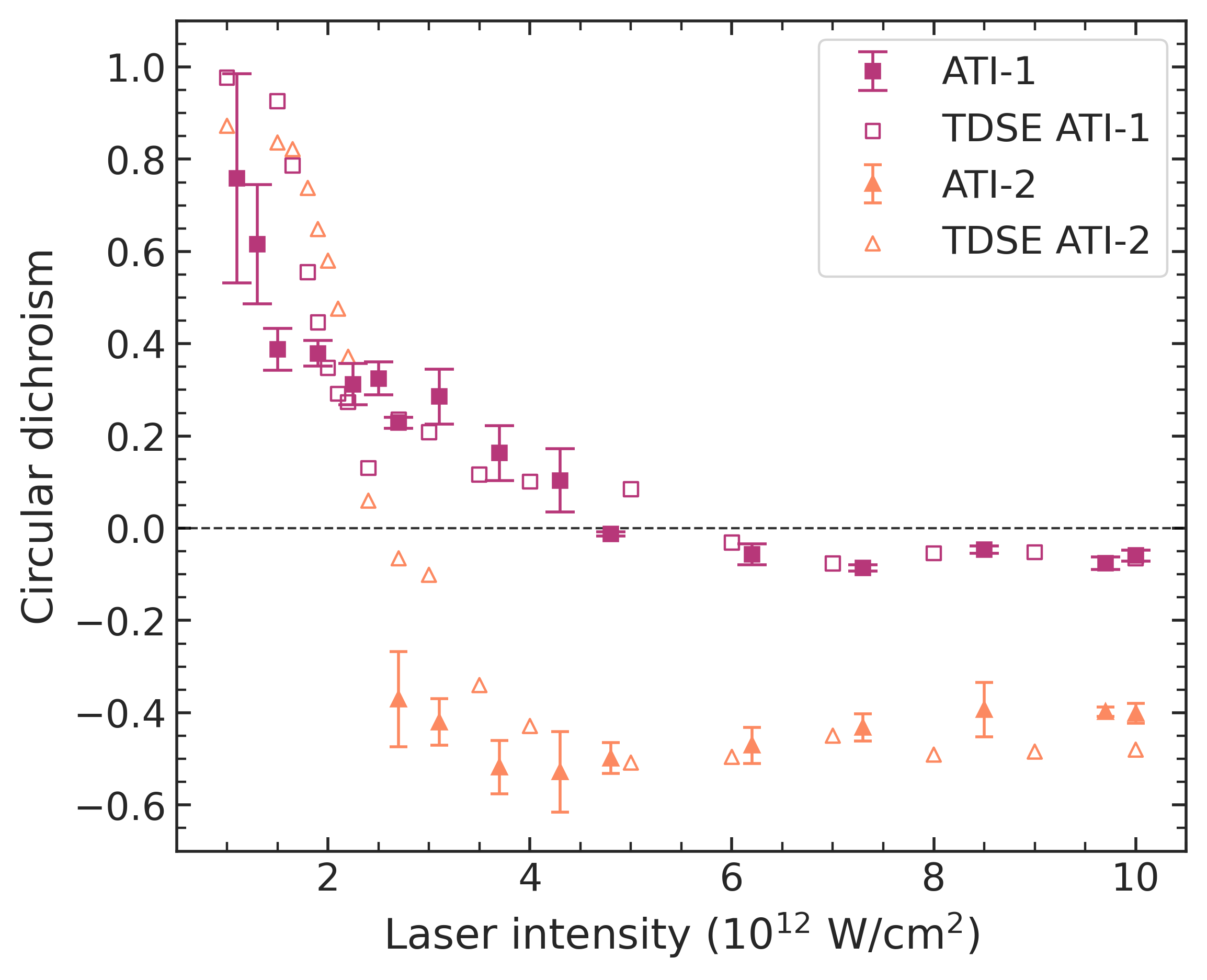}
\caption{
Experimental and theoretical CD values at different laser peak intensities for the \hbox{ATI-1} and \hbox{ATI-2} features of the photo\-electron spectrum. 
The solid symbols with error bars represent the experimental data, 
while the TDSE predictions are displayed by open symbols. 
}
\label{fig:CD_ATI}
\end{figure}

The corresponding values for the circular dichroism of the \hbox{ATI-1} and \hbox{ATI-2} lines are shown in Fig.~\ref{fig:CD_ATI}. The qualitative agreement between experiment (solid symbols) and theory (open symbols) is good.
The remaining differences are most likely due to the uncertainties associated with 
the idealized representation of the theoretical pulses while there remain 
significant uncertainties in the experimental ones, in particular regarding the focus sizes and thus the absolute intensities. 
For the \hbox{ATI-1} feature, the behavior of the CD is similar to the MPI feature, as it decreases from a value of $\approx1.0$ to slightly negative values at NIR-laser intensities exceeding $6 \times 10^{12}\,$W/cm$^2$. Interestingly, the CD of the \hbox{ATI-1} feature forms a plateau in the range  $2-3.5 \times 10^{12}\,$W/cm$^2$. This may be due to resonant transitions via high-lying Rydberg states, as channel closing appears at lower laser intensities for the co-rotating case. The theoretically predicted sharp drop of the CD at $\approx\,2.5 \times 10^{12}\,$W/cm$^2$ is not resolved experimentally. At higher laser intensities, the CD decreases again towards zero when channel closing also occurs in the counter-rotating case. \\

While the present experiment cannot provide data in the low intensity region for \hbox{ATI-2} due to a lack of signal 
(see Fig.~\ref{fig:CD_RadialProfile}), theory predicts a steep drop of the CD starting again near $1.0$ and crossing zero around $2.5 \times 10^{12}\,$W/cm$^2$. In this case, the drop in the CD continues towards large negative CD values that appear to approach $\approx\,-0.5$ at high IR intensities. The experimental data are in good agreement with the predicted strongly negative CD in the \hbox{ATI-2} peak. The lack of agreement for the low intensities may again be explained by the uncertainty in the absolute value of the experimental intensity as well as the statistical uncertainty of the data. 

At the present time, we do not have an explanation for the trends
observed in both experiment and theory for the distinctly different CDs  in the MPI, \hbox{ATI-1}, and \hbox{ATI-2} peaks. 
However, we hope that further studies on the angular distributions associated with these peaks (currently ongoing in our laboratory) will reveal additional information regarding the dominant partial-wave contributions. Such results, together with propensity rules, may shed additional light on this issue. 

\section{Conclusions and Outlook}
\label{sec:conclusion}

We obtained the CD in the photo\-ionization of oriented helium ions, specifically He$^+(3p, m=+1)$, for a variety of NIR intensities, independent of the oriented-state population. This allowed us to isolate dichroic effects and resonances and to map
their intensity dependence not only for the main photo\-electron line but also for the first and second ATI peaks. The circular dichroism of the main line assumes the maximum possible value of $+1$ at very low laser intensities.  Due to the rapid channel closing for the co-rotating case, the CD of the MPI peak drops to small positive values before rising again at even higher intensities when the four-photon channel is nearly closed for both the co- and the counter-rotating cases and only the wings of the lines contribute to the ionization signal.

The CD for the \hbox{ATI-1} and \hbox{ATI-2} lines, which require a minimum of 5 and~6 photons, respectively, is also positive at the lowest intensities for which we obtained sufficiently strong signals. The CD for the \hbox{ATI-1} line then decreases monotonically to slightly negative values with increasing NIR peak intensity, forming a small plateau between $2 - 3 \times 10^{12}\,$W/cm$^2$, while the CD for the \hbox{ATI-2} line drops from about +1 to strongly negative values around $-0.5$ at NIR peak intensities above $3 \times 10^{12}\,$W/cm$^2$.

Even though the system under investigation represents a purely Coulombic one-electron problem, there are still many interesting features to be noticed, including sub\-structures in the various peaks at particular intensities and the general trends seen in the CD for those peaks as a function of the NIR intensity. While the overall agreement between experiment and theory can be considered very satisfactory, with all principal features seen in both, simple models to explain many of these features are still missing. 

A next step for shedding light on the missing links, e.g., to the studies of atomic Li, is to vary the IR wavelength in order to further investigate the effect of the various resonances and to distinguish their role from other fundamental effects. Furthermore, angle-resolved measurements should provide additional information about the dominant angular-momentum channels and the partial-wave composition of the various pathways involved.

\section{Acknowledgements}
The authors are grateful for fruitful discussions with Thomas Pfeifer and Christian Ott from the Max Planck Institute for Nuclear Physics in Heidelberg, Germany. The authors furthermore acknowledge the invaluable support of
the technical and scientific staff of FERMI at Elettra-Sincrotrone Trieste. R.W. and M.M.\ acknowledge
funding by the Deutsche Forschungs\-gemeinschaft
(DFG) in SFB-925, ``Light induced dynamics and control of correlated quantum systems'', project No.\ 170620586. M.I., V.M., and Ph.S.\ acknowledge funding for a Peter-Paul-Ewald Fellowship from the Volks\-wagen foundation. M.I.\ acknowledges support from the DFG under Project No.\ 328961117 in SFB-1319, ``Extreme light for sensing and driving molecular chirality''. M.I., N.W.\ and M.M. were supported by the DFG Cluster of Excellence ``CUI: Advanced Imaging of Matter'', project No.\ 390715994.
The theoretical part of this work was funded by the NSF under grant Nos.\ \hbox{PHY-2110023} and \hbox{PHY-2408484} (K.B.), \hbox{No.~PHY-2012078} (N.D.), and by the ACCESS
supercomputer allocation \hbox{No.~PHY-090031}.

\section*{Data Availability Statement}
The datasets used and analysed during the current study are available from the corresponding author on reasonable request.


\begin{thebibliography}{38}%
\makeatletter
\providecommand \@ifxundefined [1]{%
 \@ifx{#1\undefined}
}%
\providecommand \@ifnum [1]{%
 \ifnum #1\expandafter \@firstoftwo
 \else \expandafter \@secondoftwo
 \fi
}%
\providecommand \@ifx [1]{%
 \ifx #1\expandafter \@firstoftwo
 \else \expandafter \@secondoftwo
 \fi
}%
\providecommand \natexlab [1]{#1}%
\providecommand \enquote  [1]{``#1''}%
\providecommand \bibnamefont  [1]{#1}%
\providecommand \bibfnamefont [1]{#1}%
\providecommand \citenamefont [1]{#1}%
\providecommand \href@noop [0]{\@secondoftwo}%
\providecommand \href [0]{\begingroup \@sanitize@url \@href}%
\providecommand \@href[1]{\@@startlink{#1}\@@href}%
\providecommand \@@href[1]{\endgroup#1\@@endlink}%
\providecommand \@sanitize@url [0]{\catcode `\\12\catcode `\$12\catcode
  `\&12\catcode `\#12\catcode `\^12\catcode `\_12\catcode `\%12\relax}%
\providecommand \@@startlink[1]{}%
\providecommand \@@endlink[0]{}%
\providecommand \url  [0]{\begingroup\@sanitize@url \@url }%
\providecommand \@url [1]{\endgroup\@href {#1}{\urlprefix }}%
\providecommand \urlprefix  [0]{URL }%
\providecommand \Eprint [0]{\href }%
\providecommand \doibase [0]{http://dx.doi.org/}%
\providecommand \selectlanguage [0]{\@gobble}%
\providecommand \bibinfo  [0]{\@secondoftwo}%
\providecommand \bibfield  [0]{\@secondoftwo}%
\providecommand \translation [1]{[#1]}%
\providecommand \BibitemOpen [0]{}%
\providecommand \bibitemStop [0]{}%
\providecommand \bibitemNoStop [0]{.\EOS\space}%
\providecommand \EOS [0]{\spacefactor3000\relax}%
\providecommand \BibitemShut  [1]{\csname bibitem#1\endcsname}%
\let\auto@bib@innerbib\@empty
\bibitem [{\citenamefont {Baier}\ \emph {et~al.}(1994)\citenamefont {Baier},
  \citenamefont {Grum-Grzhimailo},\ and\ \citenamefont
  {Kabachnik}}]{baier1994}%
  \BibitemOpen
  \bibfield  {author} {\bibinfo {author} {\bibfnamefont {S.}~\bibnamefont
  {Baier}}, \bibinfo {author} {\bibfnamefont {A.}~\bibnamefont
  {Grum-Grzhimailo}}, \ and\ \bibinfo {author} {\bibfnamefont {N.}~\bibnamefont
  {Kabachnik}},\ }\href
  {https://iopscience.iop.org/article/10.1088/0953-4075/34/11/101} {\bibfield
  {journal} {\bibinfo  {journal} {Journal of Physics B: Atomic, Molecular and
  Optical Physics}\ }\textbf {\bibinfo {volume} {27}},\ \bibinfo {pages} {3363}
  (\bibinfo {year} {1994})}\BibitemShut {NoStop}%
\bibitem [{\citenamefont {Alo{\"\i}se}\ \emph {et~al.}(2005)\citenamefont
  {Alo{\"\i}se}, \citenamefont {O’Keeffe}, \citenamefont {Cubaynes},
  \citenamefont {Meyer},\ and\ \citenamefont {Grum-Grzhimailo}}]{Aloise2005}%
  \BibitemOpen
  \bibfield  {author} {\bibinfo {author} {\bibfnamefont {S.}~\bibnamefont
  {Alo{\"\i}se}}, \bibinfo {author} {\bibfnamefont {P.}~\bibnamefont
  {O’Keeffe}}, \bibinfo {author} {\bibfnamefont {D.}~\bibnamefont
  {Cubaynes}}, \bibinfo {author} {\bibfnamefont {M.}~\bibnamefont {Meyer}}, \
  and\ \bibinfo {author} {\bibfnamefont {A.}~\bibnamefont {Grum-Grzhimailo}},\
  }\href {https://link.aps.org/doi/10.1103/PhysRevLett.94.223002} {\bibfield
  {journal} {\bibinfo  {journal} {Physical Review Letters}\ }\textbf {\bibinfo
  {volume} {94}},\ \bibinfo {pages} {223002} (\bibinfo {year}
  {2005})}\BibitemShut {NoStop}%
\bibitem [{\citenamefont {Log{\'e}}\ and\ \citenamefont
  {Boesl}(2011)}]{loge2011multiphoton}%
  \BibitemOpen
  \bibfield  {author} {\bibinfo {author} {\bibfnamefont {C.}~\bibnamefont
  {Log{\'e}}}\ and\ \bibinfo {author} {\bibfnamefont {U.}~\bibnamefont
  {Boesl}},\ }\href {https://doi.org/10.1002/cphc.201100035} {\bibfield
  {journal} {\bibinfo  {journal} {ChemPhysChem}\ }\textbf {\bibinfo {volume}
  {12}},\ \bibinfo {pages} {1940} (\bibinfo {year} {2011})}\BibitemShut
  {NoStop}%
\bibitem [{\citenamefont {Eckart}\ \emph {et~al.}(2018)\citenamefont {Eckart},
  \citenamefont {Kunitski}, \citenamefont {Richter}, \citenamefont {Hartung},
  \citenamefont {Rist}, \citenamefont {Trinter}, \citenamefont {Fehre},
  \citenamefont {Schlott}, \citenamefont {Henrichs}, \citenamefont {Schmidt}
  \emph {et~al.}}]{eckart2018ultrafast}%
  \BibitemOpen
  \bibfield  {author} {\bibinfo {author} {\bibfnamefont {S.}~\bibnamefont
  {Eckart}}, \bibinfo {author} {\bibfnamefont {M.}~\bibnamefont {Kunitski}},
  \bibinfo {author} {\bibfnamefont {M.}~\bibnamefont {Richter}}, \bibinfo
  {author} {\bibfnamefont {A.}~\bibnamefont {Hartung}}, \bibinfo {author}
  {\bibfnamefont {J.}~\bibnamefont {Rist}}, \bibinfo {author} {\bibfnamefont
  {F.}~\bibnamefont {Trinter}}, \bibinfo {author} {\bibfnamefont
  {K.}~\bibnamefont {Fehre}}, \bibinfo {author} {\bibfnamefont
  {N.}~\bibnamefont {Schlott}}, \bibinfo {author} {\bibfnamefont
  {K.}~\bibnamefont {Henrichs}}, \bibinfo {author} {\bibfnamefont {L.~P.~H.}\
  \bibnamefont {Schmidt}},  \emph {et~al.},\ }\href {\doibase
  https://doi.org/10.1038/s41567-018-0080-5} {\bibfield  {journal} {\bibinfo
  {journal} {Nature Physics}\ }\textbf {\bibinfo {volume} {14}},\ \bibinfo
  {pages} {701} (\bibinfo {year} {2018})}\BibitemShut {NoStop}%
\bibitem [{\citenamefont {De~Silva}\ \emph {et~al.}(2021)\citenamefont
  {De~Silva}, \citenamefont {Atri-Schuller}, \citenamefont {Dubey},
  \citenamefont {Acharya}, \citenamefont {Romans}, \citenamefont {Foster},
  \citenamefont {Russ}, \citenamefont {Compton}, \citenamefont {Rischbieter},
  \citenamefont {Douguet} \emph {et~al.}}]{de2021using}%
  \BibitemOpen
  \bibfield  {author} {\bibinfo {author} {\bibfnamefont {A.}~\bibnamefont
  {De~Silva}}, \bibinfo {author} {\bibfnamefont {D.}~\bibnamefont
  {Atri-Schuller}}, \bibinfo {author} {\bibfnamefont {S.}~\bibnamefont
  {Dubey}}, \bibinfo {author} {\bibfnamefont {B.}~\bibnamefont {Acharya}},
  \bibinfo {author} {\bibfnamefont {K.}~\bibnamefont {Romans}}, \bibinfo
  {author} {\bibfnamefont {K.}~\bibnamefont {Foster}}, \bibinfo {author}
  {\bibfnamefont {O.}~\bibnamefont {Russ}}, \bibinfo {author} {\bibfnamefont
  {K.}~\bibnamefont {Compton}}, \bibinfo {author} {\bibfnamefont
  {C.}~\bibnamefont {Rischbieter}}, \bibinfo {author} {\bibfnamefont
  {N.}~\bibnamefont {Douguet}},  \emph {et~al.},\ }\href
  {https://link.aps.org/doi/10.1103/PhysRevLett.126.023201} {\bibfield
  {journal} {\bibinfo  {journal} {Physical Review Letters}\ }\textbf {\bibinfo
  {volume} {126}},\ \bibinfo {pages} {023201} (\bibinfo {year}
  {2021})}\BibitemShut {NoStop}%
\bibitem [{\citenamefont {Gryzlova}\ \emph {et~al.}(2015)\citenamefont
  {Gryzlova}, \citenamefont {O’Keeffe}, \citenamefont {Cubaynes},
  \citenamefont {Garcia}, \citenamefont {Nahon}, \citenamefont
  {Grum-Grzhimailo},\ and\ \citenamefont {Meyer}}]{gryzlova2015}%
  \BibitemOpen
  \bibfield  {author} {\bibinfo {author} {\bibfnamefont {E.}~\bibnamefont
  {Gryzlova}}, \bibinfo {author} {\bibfnamefont {P.}~\bibnamefont
  {O’Keeffe}}, \bibinfo {author} {\bibfnamefont {D.}~\bibnamefont
  {Cubaynes}}, \bibinfo {author} {\bibfnamefont {G.}~\bibnamefont {Garcia}},
  \bibinfo {author} {\bibfnamefont {L.}~\bibnamefont {Nahon}}, \bibinfo
  {author} {\bibfnamefont {A.}~\bibnamefont {Grum-Grzhimailo}}, \ and\ \bibinfo
  {author} {\bibfnamefont {M.}~\bibnamefont {Meyer}},\ }\href
  {https://iopscience.iop.org/article/10.1088/1367-2630/17/4/043054/meta}
  {\bibfield  {journal} {\bibinfo  {journal} {New Journal of Physics}\ }\textbf
  {\bibinfo {volume} {17}},\ \bibinfo {pages} {043054} (\bibinfo {year}
  {2015})}\BibitemShut {NoStop}%
\bibitem [{\citenamefont {Ritchie}(1976)}]{ritchie1976theory}%
  \BibitemOpen
  \bibfield  {author} {\bibinfo {author} {\bibfnamefont {B.}~\bibnamefont
  {Ritchie}},\ }\href
  {https://journals.aps.org/pra/abstract/10.1103/PhysRevA.14.359} {\bibfield
  {journal} {\bibinfo  {journal} {Physical Review A}\ }\textbf {\bibinfo
  {volume} {14}},\ \bibinfo {pages} {359} (\bibinfo {year} {1976})}\BibitemShut
  {NoStop}%
\bibitem [{\citenamefont {B{\"o}wering}\ \emph {et~al.}(2001)\citenamefont
  {B{\"o}wering}, \citenamefont {Lischke}, \citenamefont {Schmidtke},
  \citenamefont {M{\"u}ller}, \citenamefont {Khalil},\ and\ \citenamefont
  {Heinzmann}}]{bowering2001asymmetry}%
  \BibitemOpen
  \bibfield  {author} {\bibinfo {author} {\bibfnamefont {N.}~\bibnamefont
  {B{\"o}wering}}, \bibinfo {author} {\bibfnamefont {T.}~\bibnamefont
  {Lischke}}, \bibinfo {author} {\bibfnamefont {B.}~\bibnamefont {Schmidtke}},
  \bibinfo {author} {\bibfnamefont {N.}~\bibnamefont {M{\"u}ller}}, \bibinfo
  {author} {\bibfnamefont {T.}~\bibnamefont {Khalil}}, \ and\ \bibinfo {author}
  {\bibfnamefont {U.}~\bibnamefont {Heinzmann}},\ }\href
  {https://journals.aps.org/prl/abstract/10.1103/PhysRevLett.86.1187}
  {\bibfield  {journal} {\bibinfo  {journal} {Physical Review Letters}\
  }\textbf {\bibinfo {volume} {86}},\ \bibinfo {pages} {1187} (\bibinfo {year}
  {2001})}\BibitemShut {NoStop}%
\bibitem [{\citenamefont {Nahon}\ \emph {et~al.}(2015)\citenamefont {Nahon},
  \citenamefont {Garcia},\ and\ \citenamefont {Powis}}]{nahon2015}%
  \BibitemOpen
  \bibfield  {author} {\bibinfo {author} {\bibfnamefont {L.}~\bibnamefont
  {Nahon}}, \bibinfo {author} {\bibfnamefont {G.~A.}\ \bibnamefont {Garcia}}, \
  and\ \bibinfo {author} {\bibfnamefont {I.}~\bibnamefont {Powis}},\ }\href
  {https://www.sciencedirect.com/science/article/abs/pii/S0368204815000766}
  {\bibfield  {journal} {\bibinfo  {journal} {Journal of Electron Spectroscopy
  and Related Phenomena}\ }\textbf {\bibinfo {volume} {204}},\ \bibinfo {pages}
  {322} (\bibinfo {year} {2015})}\BibitemShut {NoStop}%
\bibitem [{\citenamefont {Beaulieu}\ \emph {et~al.}(2018)\citenamefont
  {Beaulieu}, \citenamefont {Comby}, \citenamefont {Descamps}, \citenamefont
  {Fabre}, \citenamefont {Garcia}, \citenamefont {G{\'e}neaux}, \citenamefont
  {Harvey}, \citenamefont {L{\'e}gar{\'e}}, \citenamefont {Ma{\v{s}}{\'\i}n},
  \citenamefont {Nahon} \emph {et~al.}}]{beaulieu2018}%
  \BibitemOpen
  \bibfield  {author} {\bibinfo {author} {\bibfnamefont {S.}~\bibnamefont
  {Beaulieu}}, \bibinfo {author} {\bibfnamefont {A.}~\bibnamefont {Comby}},
  \bibinfo {author} {\bibfnamefont {D.}~\bibnamefont {Descamps}}, \bibinfo
  {author} {\bibfnamefont {B.}~\bibnamefont {Fabre}}, \bibinfo {author}
  {\bibfnamefont {G.}~\bibnamefont {Garcia}}, \bibinfo {author} {\bibfnamefont
  {R.}~\bibnamefont {G{\'e}neaux}}, \bibinfo {author} {\bibfnamefont
  {A.}~\bibnamefont {Harvey}}, \bibinfo {author} {\bibfnamefont
  {F.}~\bibnamefont {L{\'e}gar{\'e}}}, \bibinfo {author} {\bibfnamefont
  {Z.}~\bibnamefont {Ma{\v{s}}{\'\i}n}}, \bibinfo {author} {\bibfnamefont
  {L.}~\bibnamefont {Nahon}},  \emph {et~al.},\ }\href
  {https://www.nature.com/articles/s41567-017-0038-z} {\bibfield  {journal}
  {\bibinfo  {journal} {Nature Physics}\ }\textbf {\bibinfo {volume} {14}},\
  \bibinfo {pages} {484} (\bibinfo {year} {2018})}\BibitemShut {NoStop}%
\bibitem [{\citenamefont {Ilchen}\ \emph {et~al.}(2021)\citenamefont {Ilchen},
  \citenamefont {Schmidt}, \citenamefont {Novikovskiy}, \citenamefont
  {Hartmann}, \citenamefont {Rupprecht}, \citenamefont {Coffee}, \citenamefont
  {Ehresmann}, \citenamefont {Galler}, \citenamefont {Hartmann}, \citenamefont
  {Helml} \emph {et~al.}}]{ilchen2021site}%
  \BibitemOpen
  \bibfield  {author} {\bibinfo {author} {\bibfnamefont {M.}~\bibnamefont
  {Ilchen}}, \bibinfo {author} {\bibfnamefont {P.}~\bibnamefont {Schmidt}},
  \bibinfo {author} {\bibfnamefont {N.~M.}\ \bibnamefont {Novikovskiy}},
  \bibinfo {author} {\bibfnamefont {G.}~\bibnamefont {Hartmann}}, \bibinfo
  {author} {\bibfnamefont {P.}~\bibnamefont {Rupprecht}}, \bibinfo {author}
  {\bibfnamefont {R.~N.}\ \bibnamefont {Coffee}}, \bibinfo {author}
  {\bibfnamefont {A.}~\bibnamefont {Ehresmann}}, \bibinfo {author}
  {\bibfnamefont {A.}~\bibnamefont {Galler}}, \bibinfo {author} {\bibfnamefont
  {N.}~\bibnamefont {Hartmann}}, \bibinfo {author} {\bibfnamefont
  {W.}~\bibnamefont {Helml}},  \emph {et~al.},\ }\href
  {https://www.nature.com/articles/s42004-021-00555-6} {\bibfield  {journal}
  {\bibinfo  {journal} {Communications Chemistry}\ }\textbf {\bibinfo {volume}
  {4}},\ \bibinfo {pages} {119} (\bibinfo {year} {2021})}\BibitemShut {NoStop}%
\bibitem [{\citenamefont {Veyrinas}\ \emph {et~al.}(2019)\citenamefont
  {Veyrinas}, \citenamefont {Saquet}, \citenamefont {Marggi~Poullain},
  \citenamefont {Lebech}, \citenamefont {Houver}, \citenamefont {Lucchese},\
  and\ \citenamefont {Dowek}}]{veyrinas2019}%
  \BibitemOpen
  \bibfield  {author} {\bibinfo {author} {\bibfnamefont {K.}~\bibnamefont
  {Veyrinas}}, \bibinfo {author} {\bibfnamefont {N.}~\bibnamefont {Saquet}},
  \bibinfo {author} {\bibfnamefont {S.}~\bibnamefont {Marggi~Poullain}},
  \bibinfo {author} {\bibfnamefont {M.}~\bibnamefont {Lebech}}, \bibinfo
  {author} {\bibfnamefont {J.-C.}\ \bibnamefont {Houver}}, \bibinfo {author}
  {\bibfnamefont {R.}~\bibnamefont {Lucchese}}, \ and\ \bibinfo {author}
  {\bibfnamefont {D.}~\bibnamefont {Dowek}},\ }\href
  {https://pubs.aip.org/aip/jcp/article/151/17/174305/198366/Dissociative-photoionization-of-NO-across-a-shape}
  {\bibfield  {journal} {\bibinfo  {journal} {The Journal of Chemical Physics}\
  }\textbf {\bibinfo {volume} {151}},\ \bibinfo {pages} {174305} (\bibinfo
  {year} {2019})}\BibitemShut {NoStop}%
\bibitem [{\citenamefont {van~der Laan}\ and\ \citenamefont
  {Figueroa}(2014)}]{van2014}%
  \BibitemOpen
  \bibfield  {author} {\bibinfo {author} {\bibfnamefont {G.}~\bibnamefont
  {van~der Laan}}\ and\ \bibinfo {author} {\bibfnamefont {A.~I.}\ \bibnamefont
  {Figueroa}},\ }\href
  {https://www.sciencedirect.com/science/article/abs/pii/S0010854514000733}
  {\bibfield  {journal} {\bibinfo  {journal} {Coordination Chemistry Reviews}\
  }\textbf {\bibinfo {volume} {277}},\ \bibinfo {pages} {95} (\bibinfo {year}
  {2014})}\BibitemShut {NoStop}%
\bibitem [{\citenamefont {Barth}\ and\ \citenamefont
  {Smirnova}(2011)}]{Barth11}%
  \BibitemOpen
  \bibfield  {author} {\bibinfo {author} {\bibfnamefont {I.}~\bibnamefont
  {Barth}}\ and\ \bibinfo {author} {\bibfnamefont {O.}~\bibnamefont
  {Smirnova}},\ }\href {\doibase 10.1103/PhysRevA.84.063415} {\bibfield
  {journal} {\bibinfo  {journal} {Physical Review A}\ }\textbf {\bibinfo
  {volume} {84}},\ \bibinfo {pages} {063415} (\bibinfo {year}
  {2011})}\BibitemShut {NoStop}%
\bibitem [{\citenamefont {Mazza}\ \emph {et~al.}(2014)\citenamefont {Mazza},
  \citenamefont {Ilchen}, \citenamefont {Rafipoor}, \citenamefont {Callegari},
  \citenamefont {Finetti}, \citenamefont {Plekan}, \citenamefont {Prince},
  \citenamefont {Richter}, \citenamefont {Danailov}, \citenamefont {Demidovich}
  \emph {et~al.}}]{Mazza2014}%
  \BibitemOpen
  \bibfield  {author} {\bibinfo {author} {\bibfnamefont {T.}~\bibnamefont
  {Mazza}}, \bibinfo {author} {\bibfnamefont {M.}~\bibnamefont {Ilchen}},
  \bibinfo {author} {\bibfnamefont {A.~J.}\ \bibnamefont {Rafipoor}}, \bibinfo
  {author} {\bibfnamefont {C.}~\bibnamefont {Callegari}}, \bibinfo {author}
  {\bibfnamefont {P.}~\bibnamefont {Finetti}}, \bibinfo {author} {\bibfnamefont
  {O.}~\bibnamefont {Plekan}}, \bibinfo {author} {\bibfnamefont {K.~C.}\
  \bibnamefont {Prince}}, \bibinfo {author} {\bibfnamefont {R.}~\bibnamefont
  {Richter}}, \bibinfo {author} {\bibfnamefont {M.~B.}\ \bibnamefont
  {Danailov}}, \bibinfo {author} {\bibfnamefont {A.}~\bibnamefont
  {Demidovich}},  \emph {et~al.},\ }\href {\doibase 10.1038/ncomms4648}
  {\bibfield  {journal} {\bibinfo  {journal} {\hbox{Nature}
  \hbox{Communications}}\ }\textbf {\bibinfo {volume} {5}},\ \bibinfo {pages}
  {1} (\bibinfo {year} {2014})}\BibitemShut {NoStop}%
\bibitem [{\citenamefont {Kazansky}\ \emph {et~al.}(2011)\citenamefont
  {Kazansky}, \citenamefont {Grigorieva},\ and\ \citenamefont
  {Kabachnik}}]{kazansky2011}%
  \BibitemOpen
  \bibfield  {author} {\bibinfo {author} {\bibfnamefont {A.}~\bibnamefont
  {Kazansky}}, \bibinfo {author} {\bibfnamefont {A.}~\bibnamefont
  {Grigorieva}}, \ and\ \bibinfo {author} {\bibfnamefont {N.}~\bibnamefont
  {Kabachnik}},\ }\href
  {https://journals.aps.org/prl/abstract/10.1103/PhysRevLett.107.253002}
  {\bibfield  {journal} {\bibinfo  {journal} {Physical Review Letters}\
  }\textbf {\bibinfo {volume} {107}},\ \bibinfo {pages} {253002} (\bibinfo
  {year} {2011})}\BibitemShut {NoStop}%
\bibitem [{\citenamefont {Ilchen}\ \emph {et~al.}(2017)\citenamefont {Ilchen},
  \citenamefont {Douguet}, \citenamefont {Mazza}, \citenamefont {Rafipoor},
  \citenamefont {Callegari}, \citenamefont {Finetti}, \citenamefont {Plekan},
  \citenamefont {Prince}, \citenamefont {Demidovich}, \citenamefont {Grazioli}
  \emph {et~al.}}]{Ilchen17}%
  \BibitemOpen
  \bibfield  {author} {\bibinfo {author} {\bibfnamefont {M.}~\bibnamefont
  {Ilchen}}, \bibinfo {author} {\bibfnamefont {N.}~\bibnamefont {Douguet}},
  \bibinfo {author} {\bibfnamefont {T.}~\bibnamefont {Mazza}}, \bibinfo
  {author} {\bibfnamefont {A.~J.}\ \bibnamefont {Rafipoor}}, \bibinfo {author}
  {\bibfnamefont {C.}~\bibnamefont {Callegari}}, \bibinfo {author}
  {\bibfnamefont {P.}~\bibnamefont {Finetti}}, \bibinfo {author} {\bibfnamefont
  {O.}~\bibnamefont {Plekan}}, \bibinfo {author} {\bibfnamefont {K.~C.}\
  \bibnamefont {Prince}}, \bibinfo {author} {\bibfnamefont {A.}~\bibnamefont
  {Demidovich}}, \bibinfo {author} {\bibfnamefont {C.}~\bibnamefont
  {Grazioli}},  \emph {et~al.},\ }\href {\doibase
  10.1103/PhysRevLett.118.013002} {\bibfield  {journal} {\bibinfo  {journal}
  {Physical Review Letters}\ }\textbf {\bibinfo {volume} {118}},\ \bibinfo
  {pages} {013002} (\bibinfo {year} {2017})}\BibitemShut {NoStop}%
\bibitem [{\citenamefont {Hartmann}\ \emph {et~al.}(2016)\citenamefont
  {Hartmann}, \citenamefont {Lindahl}, \citenamefont {Knie}, \citenamefont
  {Hartmann}, \citenamefont {Lutman}, \citenamefont {MacArthur}, \citenamefont
  {Shevchuk}, \citenamefont {Buck}, \citenamefont {Galler}, \citenamefont
  {Glownia} \emph {et~al.}}]{hartmann2016circular}%
  \BibitemOpen
  \bibfield  {author} {\bibinfo {author} {\bibfnamefont {G.}~\bibnamefont
  {Hartmann}}, \bibinfo {author} {\bibfnamefont {A.}~\bibnamefont {Lindahl}},
  \bibinfo {author} {\bibfnamefont {A.}~\bibnamefont {Knie}}, \bibinfo {author}
  {\bibfnamefont {N.}~\bibnamefont {Hartmann}}, \bibinfo {author}
  {\bibfnamefont {A.}~\bibnamefont {Lutman}}, \bibinfo {author} {\bibfnamefont
  {J.}~\bibnamefont {MacArthur}}, \bibinfo {author} {\bibfnamefont
  {I.}~\bibnamefont {Shevchuk}}, \bibinfo {author} {\bibfnamefont
  {J.}~\bibnamefont {Buck}}, \bibinfo {author} {\bibfnamefont {A.}~\bibnamefont
  {Galler}}, \bibinfo {author} {\bibfnamefont {J.}~\bibnamefont {Glownia}},
  \emph {et~al.},\ }\href
  {https://pubs.aip.org/aip/rsi/article/87/8/083113/839180/Circular-dichroism-measurements-at-an-x-ray-free}
  {\bibfield  {journal} {\bibinfo  {journal} {Review of Scientific
  Instruments}\ }\textbf {\bibinfo {volume} {87}},\ \bibinfo {pages} {083113}
  (\bibinfo {year} {2016})}\BibitemShut {NoStop}%
\bibitem [{\citenamefont {Mazza}\ \emph {et~al.}(2016)\citenamefont {Mazza},
  \citenamefont {Ilchen}, \citenamefont {Rafipoor}, \citenamefont {Callegari},
  \citenamefont {Finetti}, \citenamefont {Plekan}, \citenamefont {Prince},
  \citenamefont {Richter}, \citenamefont {Demidovich}, \citenamefont {Grazioli}
  \emph {et~al.}}]{mazza2016}%
  \BibitemOpen
  \bibfield  {author} {\bibinfo {author} {\bibfnamefont {T.}~\bibnamefont
  {Mazza}}, \bibinfo {author} {\bibfnamefont {M.}~\bibnamefont {Ilchen}},
  \bibinfo {author} {\bibfnamefont {A.}~\bibnamefont {Rafipoor}}, \bibinfo
  {author} {\bibfnamefont {C.}~\bibnamefont {Callegari}}, \bibinfo {author}
  {\bibfnamefont {P.}~\bibnamefont {Finetti}}, \bibinfo {author} {\bibfnamefont
  {O.}~\bibnamefont {Plekan}}, \bibinfo {author} {\bibfnamefont
  {K.}~\bibnamefont {Prince}}, \bibinfo {author} {\bibfnamefont
  {R.}~\bibnamefont {Richter}}, \bibinfo {author} {\bibfnamefont
  {A.}~\bibnamefont {Demidovich}}, \bibinfo {author} {\bibfnamefont
  {C.}~\bibnamefont {Grazioli}},  \emph {et~al.},\ }\href
  {https://www.tandfonline.com/doi/abs/10.1080/09500340.2015.1119897}
  {\bibfield  {journal} {\bibinfo  {journal} {Journal of Modern Optics}\
  }\textbf {\bibinfo {volume} {63}},\ \bibinfo {pages} {367} (\bibinfo {year}
  {2016})}\BibitemShut {NoStop}%
\bibitem [{\citenamefont {R{\"o}rig}\ \emph {et~al.}(2023)\citenamefont
  {R{\"o}rig}, \citenamefont {Son}, \citenamefont {Mazza}, \citenamefont
  {Schmidt}, \citenamefont {Baumann}, \citenamefont {Erk}, \citenamefont
  {Ilchen}, \citenamefont {Laksman}, \citenamefont {Music}, \citenamefont
  {Pathak} \emph {et~al.}}]{rorig2023multiple}%
  \BibitemOpen
  \bibfield  {author} {\bibinfo {author} {\bibfnamefont {A.}~\bibnamefont
  {R{\"o}rig}}, \bibinfo {author} {\bibfnamefont {S.-K.}\ \bibnamefont {Son}},
  \bibinfo {author} {\bibfnamefont {T.}~\bibnamefont {Mazza}}, \bibinfo
  {author} {\bibfnamefont {P.}~\bibnamefont {Schmidt}}, \bibinfo {author}
  {\bibfnamefont {T.~M.}\ \bibnamefont {Baumann}}, \bibinfo {author}
  {\bibfnamefont {B.}~\bibnamefont {Erk}}, \bibinfo {author} {\bibfnamefont
  {M.}~\bibnamefont {Ilchen}}, \bibinfo {author} {\bibfnamefont
  {J.}~\bibnamefont {Laksman}}, \bibinfo {author} {\bibfnamefont
  {V.}~\bibnamefont {Music}}, \bibinfo {author} {\bibfnamefont
  {S.}~\bibnamefont {Pathak}},  \emph {et~al.},\ }\href
  {https://www.nature.com/articles/s41467-023-41505-1} {\bibfield  {journal}
  {\bibinfo  {journal} {Nature Communications}\ }\textbf {\bibinfo {volume}
  {14}},\ \bibinfo {pages} {5738} (\bibinfo {year} {2023})}\BibitemShut
  {NoStop}%
\bibitem [{\citenamefont {Freeman}\ and\ \citenamefont
  {Bucksbaum}(1991)}]{freeman1991}%
  \BibitemOpen
  \bibfield  {author} {\bibinfo {author} {\bibfnamefont {R.}~\bibnamefont
  {Freeman}}\ and\ \bibinfo {author} {\bibfnamefont {P.}~\bibnamefont
  {Bucksbaum}},\ }\href
  {https://iopscience.iop.org/article/10.1088/0953-4075/24/2/004} {\bibfield
  {journal} {\bibinfo  {journal} {Journal of Physics B: Atomic, Molecular and
  Optical Physics}\ }\textbf {\bibinfo {volume} {24}},\ \bibinfo {pages} {325}
  (\bibinfo {year} {1991})}\BibitemShut {NoStop}%
\bibitem [{\citenamefont {Marchenko}\ \emph {et~al.}(2010)\citenamefont
  {Marchenko}, \citenamefont {Muller}, \citenamefont {Schafer},\ and\
  \citenamefont {Vrakking}}]{marchenko2010}%
  \BibitemOpen
  \bibfield  {author} {\bibinfo {author} {\bibfnamefont {T.}~\bibnamefont
  {Marchenko}}, \bibinfo {author} {\bibfnamefont {H.}~\bibnamefont {Muller}},
  \bibinfo {author} {\bibfnamefont {K.}~\bibnamefont {Schafer}}, \ and\
  \bibinfo {author} {\bibfnamefont {M.}~\bibnamefont {Vrakking}},\ }\href
  {https://iopscience.iop.org/article/10.1088/0953-4075/43/18/185001}
  {\bibfield  {journal} {\bibinfo  {journal} {Journal of Physics B: Atomic,
  Molecular and Optical Physics}\ }\textbf {\bibinfo {volume} {43}},\ \bibinfo
  {pages} {185001} (\bibinfo {year} {2010})}\BibitemShut {NoStop}%
\bibitem [{\citenamefont {Allaria}\ \emph {et~al.}(2014)\citenamefont
  {Allaria}, \citenamefont {Diviacco}, \citenamefont {Callegari}, \citenamefont
  {Finetti}, \citenamefont {Mahieu}, \citenamefont {Viefhaus}, \citenamefont
  {Zangrando}, \citenamefont {De~Ninno}, \citenamefont {Lambert}, \citenamefont
  {Ferrari} \emph {et~al.}}]{fermi_allaria}%
  \BibitemOpen
  \bibfield  {author} {\bibinfo {author} {\bibfnamefont {E.}~\bibnamefont
  {Allaria}}, \bibinfo {author} {\bibfnamefont {B.}~\bibnamefont {Diviacco}},
  \bibinfo {author} {\bibfnamefont {C.}~\bibnamefont {Callegari}}, \bibinfo
  {author} {\bibfnamefont {P.}~\bibnamefont {Finetti}}, \bibinfo {author}
  {\bibfnamefont {B.}~\bibnamefont {Mahieu}}, \bibinfo {author} {\bibfnamefont
  {J.}~\bibnamefont {Viefhaus}}, \bibinfo {author} {\bibfnamefont
  {M.}~\bibnamefont {Zangrando}}, \bibinfo {author} {\bibfnamefont
  {G.}~\bibnamefont {De~Ninno}}, \bibinfo {author} {\bibfnamefont
  {G.}~\bibnamefont {Lambert}}, \bibinfo {author} {\bibfnamefont
  {E.}~\bibnamefont {Ferrari}},  \emph {et~al.},\ }\href {\doibase
  10.1103/PhysRevX.4.041040} {\bibfield  {journal} {\bibinfo  {journal}
  {Physical Review X}\ }\textbf {\bibinfo {volume} {4}},\ \bibinfo {pages}
  {041040} (\bibinfo {year} {2014})}\BibitemShut {NoStop}%
\bibitem [{\citenamefont {Bauer}\ \emph {et~al.}(2014)\citenamefont {Bauer},
  \citenamefont {Mota-Furtado}, \citenamefont {O'Mahony}, \citenamefont
  {Piraux},\ and\ \citenamefont {Warda}}]{Bauer14}%
  \BibitemOpen
  \bibfield  {author} {\bibinfo {author} {\bibfnamefont {J.~H.}\ \bibnamefont
  {Bauer}}, \bibinfo {author} {\bibfnamefont {F.}~\bibnamefont {Mota-Furtado}},
  \bibinfo {author} {\bibfnamefont {P.~F.}\ \bibnamefont {O'Mahony}}, \bibinfo
  {author} {\bibfnamefont {B.}~\bibnamefont {Piraux}}, \ and\ \bibinfo {author}
  {\bibfnamefont {K.}~\bibnamefont {Warda}},\ }\href {\doibase
  10.1103/PhysRevA.90.063402} {\bibfield  {journal} {\bibinfo  {journal}
  {Physical Review A}\ }\textbf {\bibinfo {volume} {90}},\ \bibinfo {pages}
  {063402} (\bibinfo {year} {2014})}\BibitemShut {NoStop}%
\bibitem [{\citenamefont {Lyamayev}\ \emph {et~al.}(2013)\citenamefont
  {Lyamayev}, \citenamefont {Ovcharenko}, \citenamefont {Katzy}, \citenamefont
  {Devetta}, \citenamefont {Bruder}, \citenamefont {LaForge}, \citenamefont
  {Mudrich}, \citenamefont {Person}, \citenamefont {Stienkemeier},
  \citenamefont {Krikunova} \emph {et~al.}}]{Lyamayev_2013}%
  \BibitemOpen
  \bibfield  {author} {\bibinfo {author} {\bibfnamefont {V.}~\bibnamefont
  {Lyamayev}}, \bibinfo {author} {\bibfnamefont {Y.}~\bibnamefont
  {Ovcharenko}}, \bibinfo {author} {\bibfnamefont {R.}~\bibnamefont {Katzy}},
  \bibinfo {author} {\bibfnamefont {M.}~\bibnamefont {Devetta}}, \bibinfo
  {author} {\bibfnamefont {L.}~\bibnamefont {Bruder}}, \bibinfo {author}
  {\bibfnamefont {A.}~\bibnamefont {LaForge}}, \bibinfo {author} {\bibfnamefont
  {M.}~\bibnamefont {Mudrich}}, \bibinfo {author} {\bibfnamefont
  {U.}~\bibnamefont {Person}}, \bibinfo {author} {\bibfnamefont
  {F.}~\bibnamefont {Stienkemeier}}, \bibinfo {author} {\bibfnamefont
  {M.}~\bibnamefont {Krikunova}},  \emph {et~al.},\ }\href {\doibase
  10.1088/0953-4075/46/16/164007} {\bibfield  {journal} {\bibinfo  {journal}
  {Journal of Physics B: Atomic, Molecular and Optical Physics}\ }\textbf
  {\bibinfo {volume} {46}},\ \bibinfo {pages} {164007} (\bibinfo {year}
  {2013})}\BibitemShut {NoStop}%
\bibitem [{\citenamefont {Allaria}\ \emph {et~al.}(2012)\citenamefont
  {Allaria}, \citenamefont {Appio}, \citenamefont {Badano}, \citenamefont
  {Barletta}, \citenamefont {Bassanese}, \citenamefont {Biedron}, \citenamefont
  {Borga}, \citenamefont {Busetto}, \citenamefont {Castronovo}, \citenamefont
  {Cinquegrana} \emph {et~al.}}]{allaria2012}%
  \BibitemOpen
  \bibfield  {author} {\bibinfo {author} {\bibfnamefont {E.}~\bibnamefont
  {Allaria}}, \bibinfo {author} {\bibfnamefont {R.}~\bibnamefont {Appio}},
  \bibinfo {author} {\bibfnamefont {L.}~\bibnamefont {Badano}}, \bibinfo
  {author} {\bibfnamefont {W.}~\bibnamefont {Barletta}}, \bibinfo {author}
  {\bibfnamefont {S.}~\bibnamefont {Bassanese}}, \bibinfo {author}
  {\bibfnamefont {S.}~\bibnamefont {Biedron}}, \bibinfo {author} {\bibfnamefont
  {A.}~\bibnamefont {Borga}}, \bibinfo {author} {\bibfnamefont
  {E.}~\bibnamefont {Busetto}}, \bibinfo {author} {\bibfnamefont
  {D.}~\bibnamefont {Castronovo}}, \bibinfo {author} {\bibfnamefont
  {P.}~\bibnamefont {Cinquegrana}},  \emph {et~al.},\ }\href
  {https://doi.org/10.1038/nphoton.2012.233} {\bibfield  {journal} {\bibinfo
  {journal} {Nature Photonics}\ }\textbf {\bibinfo {volume} {6}},\ \bibinfo
  {pages} {699} (\bibinfo {year} {2012})}\BibitemShut {NoStop}%
\bibitem [{\citenamefont {Kramida}\ \emph {et~al.}(2022)\citenamefont
  {Kramida}, \citenamefont {Ralchenko}, \citenamefont {Reader},\ and\
  \citenamefont {{NIST ASD Team}}}]{NIST}%
  \BibitemOpen
  \bibfield  {author} {\bibinfo {author} {\bibfnamefont {A.~E.}\ \bibnamefont
  {Kramida}}, \bibinfo {author} {\bibfnamefont {Y.}~\bibnamefont {Ralchenko}},
  \bibinfo {author} {\bibfnamefont {J.}~\bibnamefont {Reader}}, \ and\ \bibinfo
  {author} {\bibnamefont {{NIST ASD Team}}},\ }\href
  {https://www.nist.gov/pml/atomic-spectra-database} {\enquote {\bibinfo
  {title} {{NIST} {A}tomic {S}pectra {D}atabase, (version 5.10)},}\ } (\bibinfo
  {year} {2022})\BibitemShut {NoStop}%
\bibitem [{\citenamefont {Bethe}\ and\ \citenamefont
  {Salpeter}(2013)}]{bethe2013quantum}%
  \BibitemOpen
  \bibfield  {author} {\bibinfo {author} {\bibfnamefont {H.}~\bibnamefont
  {Bethe}}\ and\ \bibinfo {author} {\bibfnamefont {E.}~\bibnamefont
  {Salpeter}},\ }\href {https://books.google.de/books?id=nxz2CAAAQBAJ} {\emph
  {\bibinfo {title} {Quantum Mechanics of One- and Two-Electron Atoms}}}\
  (\bibinfo  {publisher} {Springer Berlin Heidelberg},\ \bibinfo {year}
  {2013})\BibitemShut {NoStop}%
\bibitem [{\citenamefont {Douguet}\ \emph {et~al.}(2024)\citenamefont
  {Douguet}, \citenamefont {Guchkov}, \citenamefont {Bartschat},\ and\
  \citenamefont {Santos}}]{atoms12070034}%
  \BibitemOpen
  \bibfield  {author} {\bibinfo {author} {\bibfnamefont {N.}~\bibnamefont
  {Douguet}}, \bibinfo {author} {\bibfnamefont {M.}~\bibnamefont {Guchkov}},
  \bibinfo {author} {\bibfnamefont {K.}~\bibnamefont {Bartschat}}, \ and\
  \bibinfo {author} {\bibfnamefont {S.~F.~d.}\ \bibnamefont {Santos}},\ }\href
  {\doibase 10.3390/atoms12070034} {\bibfield  {journal} {\bibinfo  {journal}
  {Atoms}\ }\textbf {\bibinfo {volume} {12}} (\bibinfo {year} {2024}),\
  10.3390/atoms12070034}\BibitemShut {NoStop}%
\bibitem [{\citenamefont {Cormier}\ and\ \citenamefont
  {Lambropoulos}(1996)}]{CorLam1996}%
  \BibitemOpen
  \bibfield  {author} {\bibinfo {author} {\bibfnamefont {E.}~\bibnamefont
  {Cormier}}\ and\ \bibinfo {author} {\bibfnamefont {P.}~\bibnamefont
  {Lambropoulos}},\ }\href {\doibase 10.1088/0953-4075/29/9/013} {\bibfield
  {journal} {\bibinfo  {journal} {Journal of Physics B: Atomic, Molecular and
  Optical Physics}\ }\textbf {\bibinfo {volume} {29}},\ \bibinfo {pages} {1667}
  (\bibinfo {year} {1996})}\BibitemShut {NoStop}%
\bibitem [{\citenamefont {Grum-Grzhimailo}\ \emph {et~al.}(2010)\citenamefont
  {Grum-Grzhimailo}, \citenamefont {Abeln}, \citenamefont {Bartschat},
  \citenamefont {\hbox{Weflen}},\ and\ \citenamefont {Urness}}]{Grum10}%
  \BibitemOpen
  \bibfield  {author} {\bibinfo {author} {\bibfnamefont {A.~N.}\ \bibnamefont
  {Grum-Grzhimailo}}, \bibinfo {author} {\bibfnamefont {B.}~\bibnamefont
  {Abeln}}, \bibinfo {author} {\bibfnamefont {K.}~\bibnamefont {Bartschat}},
  \bibinfo {author} {\bibfnamefont {D.}~\bibnamefont {\hbox{Weflen}}}, \ and\
  \bibinfo {author} {\bibfnamefont {T.}~\bibnamefont {Urness}},\ }\href
  {\doibase 10.1103/PhysRevA.81.043408} {\bibfield  {journal} {\bibinfo
  {journal} {Physical Review A}\ }\textbf {\bibinfo {volume} {81}},\ \bibinfo
  {pages} {043408} (\bibinfo {year} {2010})}\BibitemShut {NoStop}%
\bibitem [{\citenamefont {Ivanov}\ \emph {et~al.}(2014)\citenamefont {Ivanov},
  \citenamefont {Kheifets}, \citenamefont {Bartschat}, \citenamefont
  {\hbox{Emmons}}, \citenamefont {Buczek}, \citenamefont {Gryzlova},\ and\
  \citenamefont {Grum-Grzhimailo}}]{PhysRevA.90.043401}%
  \BibitemOpen
  \bibfield  {author} {\bibinfo {author} {\bibfnamefont {I.~A.}\ \bibnamefont
  {Ivanov}}, \bibinfo {author} {\bibfnamefont {A.~S.}\ \bibnamefont
  {Kheifets}}, \bibinfo {author} {\bibfnamefont {K.}~\bibnamefont {Bartschat}},
  \bibinfo {author} {\bibfnamefont {J.}~\bibnamefont {\hbox{Emmons}}}, \bibinfo
  {author} {\bibfnamefont {S.~M.}\ \bibnamefont {Buczek}}, \bibinfo {author}
  {\bibfnamefont {E.~V.}\ \bibnamefont {Gryzlova}}, \ and\ \bibinfo {author}
  {\bibfnamefont {A.~N.}\ \bibnamefont {Grum-Grzhimailo}},\ }\href {\doibase
  10.1103/PhysRevA.90.043401} {\bibfield  {journal} {\bibinfo  {journal}
  {Physical Review A}\ }\textbf {\bibinfo {volume} {90}},\ \bibinfo {pages}
  {043401} (\bibinfo {year} {2014})}\BibitemShut {NoStop}%
\bibitem [{\citenamefont {Grum-Grzhimailo}\ \emph {et~al.}(2019)\citenamefont
  {Grum-Grzhimailo}, \citenamefont {Douguet}, \citenamefont {Meyer},\ and\
  \citenamefont {Bartschat}}]{PhysRevA.100.033404}%
  \BibitemOpen
  \bibfield  {author} {\bibinfo {author} {\bibfnamefont {A.~N.}\ \bibnamefont
  {Grum-Grzhimailo}}, \bibinfo {author} {\bibfnamefont {N.}~\bibnamefont
  {Douguet}}, \bibinfo {author} {\bibfnamefont {M.}~\bibnamefont {Meyer}}, \
  and\ \bibinfo {author} {\bibfnamefont {K.}~\bibnamefont {Bartschat}},\ }\href
  {\doibase 10.1103/PhysRevA.100.033404} {\bibfield  {journal} {\bibinfo
  {journal} {Physical Review A}\ }\textbf {\bibinfo {volume} {100}},\ \bibinfo
  {pages} {033404} (\bibinfo {year} {2019})}\BibitemShut {NoStop}%
\bibitem [{\citenamefont {Acharya}\ \emph {et~al.}(2022)\citenamefont
  {Acharya}, \citenamefont {Dubey}, \citenamefont {Romans}, \citenamefont
  {De~Silva}, \citenamefont {Foster}, \citenamefont {Russ}, \citenamefont
  {Bartschat}, \citenamefont {Douguet},\ and\ \citenamefont
  {Fischer}}]{acharya2022two}%
  \BibitemOpen
  \bibfield  {author} {\bibinfo {author} {\bibfnamefont {B.}~\bibnamefont
  {Acharya}}, \bibinfo {author} {\bibfnamefont {S.}~\bibnamefont {Dubey}},
  \bibinfo {author} {\bibfnamefont {K.}~\bibnamefont {Romans}}, \bibinfo
  {author} {\bibfnamefont {A.}~\bibnamefont {De~Silva}}, \bibinfo {author}
  {\bibfnamefont {K.}~\bibnamefont {Foster}}, \bibinfo {author} {\bibfnamefont
  {O.}~\bibnamefont {Russ}}, \bibinfo {author} {\bibfnamefont {K.}~\bibnamefont
  {Bartschat}}, \bibinfo {author} {\bibfnamefont {N.}~\bibnamefont {Douguet}},
  \ and\ \bibinfo {author} {\bibfnamefont {D.}~\bibnamefont {Fischer}},\ }\href
  {https://link.aps.org/doi/10.1103/PhysRevA.106.023113} {\bibfield  {journal}
  {\bibinfo  {journal} {Physical Review A}\ }\textbf {\bibinfo {volume}
  {106}},\ \bibinfo {pages} {023113} (\bibinfo {year} {2022})}\BibitemShut
  {NoStop}%
\bibitem [{\citenamefont {Freeman}\ \emph {et~al.}(1987)\citenamefont
  {Freeman}, \citenamefont {Bucksbaum}, \citenamefont {Milchberg},
  \citenamefont {Darack}, \citenamefont {Schumacher},\ and\ \citenamefont
  {Geusic}}]{PhysRevLett.59.1092}%
  \BibitemOpen
  \bibfield  {author} {\bibinfo {author} {\bibfnamefont {R.~R.}\ \bibnamefont
  {Freeman}}, \bibinfo {author} {\bibfnamefont {P.~H.}\ \bibnamefont
  {Bucksbaum}}, \bibinfo {author} {\bibfnamefont {H.}~\bibnamefont
  {Milchberg}}, \bibinfo {author} {\bibfnamefont {S.}~\bibnamefont {Darack}},
  \bibinfo {author} {\bibfnamefont {D.}~\bibnamefont {Schumacher}}, \ and\
  \bibinfo {author} {\bibfnamefont {M.~E.}\ \bibnamefont {Geusic}},\ }\href
  {\doibase 10.1103/PhysRevLett.59.1092} {\bibfield  {journal} {\bibinfo
  {journal} {Physical Review Letters}\ }\textbf {\bibinfo {volume} {59}},\
  \bibinfo {pages} {1092} (\bibinfo {year} {1987})}\BibitemShut {NoStop}%
\bibitem [{\citenamefont {Hartung}\ \emph {et~al.}(2021)\citenamefont
  {Hartung}, \citenamefont {Brennecke}, \citenamefont {Lin}, \citenamefont
  {Trabert}, \citenamefont {Fehre}, \citenamefont {Rist}, \citenamefont
  {Sch{\"o}ffler}, \citenamefont {Jahnke}, \citenamefont {Schmidt},
  \citenamefont {Kunitski} \emph {et~al.}}]{hartung2021electric}%
  \BibitemOpen
  \bibfield  {author} {\bibinfo {author} {\bibfnamefont {A.}~\bibnamefont
  {Hartung}}, \bibinfo {author} {\bibfnamefont {S.}~\bibnamefont {Brennecke}},
  \bibinfo {author} {\bibfnamefont {K.}~\bibnamefont {Lin}}, \bibinfo {author}
  {\bibfnamefont {D.}~\bibnamefont {Trabert}}, \bibinfo {author} {\bibfnamefont
  {K.}~\bibnamefont {Fehre}}, \bibinfo {author} {\bibfnamefont
  {J.}~\bibnamefont {Rist}}, \bibinfo {author} {\bibfnamefont {M.}~\bibnamefont
  {Sch{\"o}ffler}}, \bibinfo {author} {\bibfnamefont {T.}~\bibnamefont
  {Jahnke}}, \bibinfo {author} {\bibfnamefont {L.~P.~H.}\ \bibnamefont
  {Schmidt}}, \bibinfo {author} {\bibfnamefont {M.}~\bibnamefont {Kunitski}},
  \emph {et~al.},\ }\href
  {https://link.aps.org/doi/10.1103/PhysRevLett.126.053202} {\bibfield
  {journal} {\bibinfo  {journal} {Physical Review Letters}\ }\textbf {\bibinfo
  {volume} {126}},\ \bibinfo {pages} {053202} (\bibinfo {year}
  {2021})}\BibitemShut {NoStop}%
\bibitem [{\citenamefont {Lin}\ \emph {et~al.}(2022)\citenamefont {Lin},
  \citenamefont {Brennecke}, \citenamefont {Ni}, \citenamefont {Chen},
  \citenamefont {Hartung}, \citenamefont {Trabert}, \citenamefont {Fehre},
  \citenamefont {Rist}, \citenamefont {Tong}, \citenamefont {Burgd{\"o}rfer}
  \emph {et~al.}}]{lin2022magnetic}%
  \BibitemOpen
  \bibfield  {author} {\bibinfo {author} {\bibfnamefont {K.}~\bibnamefont
  {Lin}}, \bibinfo {author} {\bibfnamefont {S.}~\bibnamefont {Brennecke}},
  \bibinfo {author} {\bibfnamefont {H.}~\bibnamefont {Ni}}, \bibinfo {author}
  {\bibfnamefont {X.}~\bibnamefont {Chen}}, \bibinfo {author} {\bibfnamefont
  {A.}~\bibnamefont {Hartung}}, \bibinfo {author} {\bibfnamefont
  {D.}~\bibnamefont {Trabert}}, \bibinfo {author} {\bibfnamefont
  {K.}~\bibnamefont {Fehre}}, \bibinfo {author} {\bibfnamefont
  {J.}~\bibnamefont {Rist}}, \bibinfo {author} {\bibfnamefont {X.-M.}\
  \bibnamefont {Tong}}, \bibinfo {author} {\bibfnamefont {J.}~\bibnamefont
  {Burgd{\"o}rfer}},  \emph {et~al.},\ }\href
  {https://link.aps.org/doi/10.1103/PhysRevLett.128.023201} {\bibfield
  {journal} {\bibinfo  {journal} {Physical Review Letters}\ }\textbf {\bibinfo
  {volume} {128}},\ \bibinfo {pages} {023201} (\bibinfo {year}
  {2022})}\BibitemShut {NoStop}%
\bibitem [{\citenamefont {Wang}\ \emph {et~al.}(2020)\citenamefont {Wang},
  \citenamefont {Jiang}, \citenamefont {Tian},\ and\ \citenamefont
  {Sun}}]{Wang2020}%
  \BibitemOpen
  \bibfield  {author} {\bibinfo {author} {\bibfnamefont {S.}~\bibnamefont
  {Wang}}, \bibinfo {author} {\bibfnamefont {W.-C.}\ \bibnamefont {Jiang}},
  \bibinfo {author} {\bibfnamefont {X.-Q.}\ \bibnamefont {Tian}}, \ and\
  \bibinfo {author} {\bibfnamefont {H.-B.}\ \bibnamefont {Sun}},\ }\href
  {\doibase 10.1103/PhysRevA.101.053417} {\bibfield  {journal} {\bibinfo
  {journal} {Physical Review A}\ }\textbf {\bibinfo {volume} {101}},\ \bibinfo
  {pages} {053417} (\bibinfo {year} {2020})}\BibitemShut {NoStop}%
\end{thebibliography}

%

\end{document}